\documentclass[fleqn,usenatbib]{mnras}
\usepackage{newtxtext,newtxmath}
\usepackage[T1]{fontenc}
\usepackage{xcolor,colortbl}

\DeclareRobustCommand{\VAN}[3]{#2}
\let\VANthebibliography\thebibliography
\def\thebibliography{\DeclareRobustCommand{\VAN}[3]{##3}\VANthebibliography}

\newcommand\T{\rule{0pt}{2.6ex}}       
\newcommand\B{\rule[-1.2ex]{0pt}{0pt}} 

\usepackage{graphicx}
\usepackage{upgreek}
\usepackage{stfloats}
\usepackage{amsmath}
\newcommand{\bs}[1]{\boldsymbol{#1}}

\title[The impact of magnetic fields on galaxy mergers. I]{The impact of magnetic fields on cosmological galaxy mergers. \\I: Reshaping gas and stellar discs}

\author[Whittingham et al.]{
Joseph Whittingham$^{1,2}\thanks{E-mail: jwhittingham@aip.de (AIP)}$, 
Martin Sparre$^{2,1}$, 
Christoph Pfrommer$^{1}$, 
R{\"u}diger Pakmor$^{3}$
\\
$^{1}$Leibniz-Institute for Astrophysics Potsdam (AIP), An der Sternwarte 16, 14482 Potsdam, Germany\\
$^2$Institut f\"ur Physik und Astronomie, Universit\"at Potsdam, Karl-Liebknecht-Str.\,24/25, 14476 Potsdam, Germany\\
$^{3}$Max Planck Institute for Astrophysics, Karl-Schwarzschild-Str. 1, 85741 Garching, Germany}

\date{Accepted XXX. Received YYY; in original form ZZZ}

\pubyear{2020}

\begin{document}
\label{firstpage}
\pagerange{\pageref{firstpage}--\pageref{lastpage}}
\maketitle
 
\begin{abstract}
Mergers play an important role in galaxy evolution. In particular, major mergers are able to have a transformative effect on galaxy morphology. In this paper, we investigate the role of magnetic fields in gas-rich major mergers. To this end, we run a series of high-resolution magnetohydrodynamic (MHD) zoom-in simulations with the moving-mesh code \textsc{arepo} and compare the outcome with hydrodynamic simulations run from the same initial conditions. This is the first time that the effect of magnetic fields in major mergers has been investigated in a cosmologically-consistent manner. In contrast to previous non-cosmological simulations, we find that the inclusion of magnetic fields has a substantial impact on the production of the merger remnant. Whilst magnetic fields do not strongly affect global properties, such as the star formation history, they are able to significantly influence structural properties. Indeed, MHD simulations consistently form remnants with extended discs and well-developed spiral structure, whilst hydrodynamic simulations form more compact remnants that display distinctive ring morphology. We support this work with a resolution study and show that whilst global properties are broadly converged across resolution and physics models, morphological differences only develop given sufficient resolution. We argue that this is due to the more efficient excitement of a small-scale dynamo in higher resolution simulations, resulting in a more strongly amplified field that is better able to influence gas dynamics.
\end{abstract}


\section{Introduction} \label{sec:introduction}

Radio synchrotron observations, amongst other evidence, show that spiral galaxies in the local Universe are permeated by magnetic fields \citep{beck1985}. On the galactic scale these fields are remarkably ordered and have typical strengths of a few $\upmu$G at solar radii \citep{beck2011}, rising to $50-100\;\upmu$G in nuclear starburst regions \citep{heesen2011, adebahr2013}. At these strengths, the magnetic field contributes significantly to the total pressure in the interstellar medium (ISM). Indeed, it is generally believed that the magnetic energy density in the ISM is in rough equipartition with the thermal gas, turbulent, and cosmic ray energy densities \citep{beck1996, beck2015}. Magnetic fields are therefore expected to be dynamically important for late-type galaxies today, helping to balance the disc against gravitation and directly influencing the flow of gas. On top of this, galactic magnetic fields are able to have a significant indirect impact; they can conduct collisionless particle species, such as cosmic rays, along their flux tubes \citep{Zweibel2017,thomas2020} thereby mediating the direction of momentum and energy transfer from cosmic rays to the thermal gas. This transfer enables dynamical feedback that is able to drive galactic winds, which affects galaxy formation and evolution \citep{Hanasz2013,Pakmor2016II,Ruszkowski2017,Jacob2018}.

Whilst magnetic fields may be influential at the present epoch, their direct \textit{long-term} effect on galaxy evolution is, however, still debated. It has been argued that a sufficiently strong primordial field could lead to reduced disc sizes \citep{martin-alvarez2020} and suppressed star formation rates \citep{marinacci2016}. The seed field strength required to produce these effects, though, is several magnitudes higher than that plausibly generated by battery processes in ionisation fronts or cosmological shocks \citep{kulsrad2008}. In cosmological simulations that used weaker initial seed fields, the galactic magnetic field was not able to have a significant evolutionary impact as it was amplified in a dynamo close to equipartition strengths, at which point it is quenched by magnetic tension and thus does not become strong enough to have a significant dynamical backreaction \citep{pakmor2017}. Moreover, in some simulations the field was unable to reach equipartition at all \citep{hopkins2020}, preventing it from influencing galactic development except indirectly via its impact on anisotropic transport processes. The investigations published thus far have, however, focused almost entirely on galaxies with quiescent merger histories.

In a $\Lambda$CDM cosmology -- i.e. cold dark matter with a cosmological constant, $\Lambda$ -- structure forms hierarchically and consequently few if any galaxies will remain completely untouched by interactions during their lifetime. This is especially true of more massive galaxies (with stellar mass M$_\star \gtrsim 5 \times 10^{10} \; $M$_\odot$), where mergers are suggested to be the main drivers of growth at redshifts $z\lesssim1$ \citep{bell2006, tacchella2019}. Such mergers give rise to a rapidly varying gravitational potential, which can have dramatic consequences for the galactic components and their kinematics. Being collisional, the gas component is particularly sensitive to such interactions. Indeed, it has long been recognized that mergers can draw fresh gas deep into the galaxy \citep{toomre1972, barnes1996, moreno2020}, diluting the metallicity of the existing gas \citep{scudder2012, torrey2012, bustamante2018, thorp2019, bustamante2020}, triggering starbursts \citep{farrah2003, cox2008, teyssier2010, hayward2014, luo2014} and increasing the black hole accretion rate \citep{springel2005a, sijacki2007, gabor2016}.  If the resultant feedback from these events is too strong, the gas may then be expelled from the galaxy almost entirely, quenching further star formation and transforming the galaxy into a so-called \textit{red and dead} galaxy -- an evolutionary track codified in \citet{hopkins2008}. This evolution is not necessarily pre-determined though; in fact, an increasing body of evidence shows that gas-rich mergers can instead support the growth of a stellar disc post-merger \citep{sparre2017, rodriguez-gomez2017, hani2020}. In either case, it is clear that the gas dynamics play a crucial role in the outcome.

Whilst the impact of mergers on the gas component (and vice versa) has been well-appreciated, the potential role that magnetic fields play is often neglected. However, these elements are not easily separated. Outside the densest parts of molecular clouds, the gas in galaxies is sufficiently ionised such that even the neutral component is intimately coupled to the magnetic field \citep{ferriere2001}. This has clear consequences during a merger; as gas is sheared and compressed, so too will field lines be brought closer together. Similarly, the injection of turbulence during a tidal interaction should support the development of a small-scale dynamo \citep{arshakian2009}. These processes will act to amplify the field, and help to explain why interacting galaxies show lower field regularity (the ratio of regular to random field components) and higher field strengths than non-interacting galaxies \citep{drzazga2011}. Given sufficient and rapid enough amplification, it is possible that the galactic magnetic field could reach equipartition or even become locally dominant during the merger. This would have significant ramifications for the gas dynamics and subsequent star formation.

Such a scenario has not yet been rigorously tested. Indeed, merger simulations have traditionally been performed using pure hydrodynamics only. This owes both to the well-known technical difficulties in discretising the equations of MHD whilst sufficiently maintaining the $\bs{\nabla} \bs{\cdot} \bs{B} = 0$ constraint in dynamic environments, and to the already considerable computational cost involved in sufficiently resolving the galactic interior. To increase resolution, the few MHD simulations that have focused on mergers have been run exclusively with idealised set-ups \citep{kotarba2010, kotarba2011, geng2012, moss2014, rodenbeck2016}. Whilst these can be helpful to probe the physics involved, such simulations necessarily require the somewhat arbitrary choice of a range of free parameters. This can result in cosmologically-inconsistent mass infall, tidal fields, and orbital parameters for the participating galaxies. This is problematic as such parameters have been shown to play a pivotal role in the production of the merger remnant \citep[e.g.][]{naab2003}.

The choice of free parameters may be reduced by running fully-cosmological simulations. However, to be cosmologically-consistent, the galactic environment must be resolved for several tens of Mpc. In contrast, resolving the turbulence in the ISM that drives the small-scale dynamo is expected to require close to parsec resolution \citep{dubois2010, renaud2014}. Clearly, resolving these scales at the same resolution is computationally infeasible. In this work, we attempt to reconcile the differences between these scales by running cosmological MHD `zoom-in' simulations, which focus their computational power on the object of interest and resolve the surrounding environment more coarsely. Such simulations are able to include large-scale cosmological effects, whilst resolving baryonic processes below the kpc scale. Several MHD zoom-in simulations of this kind have now been run \cite[e.g.][]{rieder2017, pakmor2017, hopkins2020, libeskind2020}. However, as already stated, until now these have all focused on relatively isolated galaxies, with any analysis on mergers being purely incidental \citep[e.g.][]{pakmor2014, martin-alvarez2018}. In this paper, we rectify this by presenting a series of high-resolution cosmological MHD zoom-in simulations of major mergers. By comparing the outcome of these simulations to hydrodynamic simulations run from the same initial conditions, we evaluate the impact of magnetic fields on galaxy mergers in a cosmologically-consistent manner for the first time.

The paper is organised as follows: in Section~\ref{sec:methodology} we describe our methodology and present the simulations. In Section~\ref{sec:analysis}, we present our analysis of the simulations, including: the impact of MHD on global properties (Section~\ref{subsec:MHD-global-properties}), the impact of MHD on structural properties (Section~\ref{subsec:MHD-disc-sizes}), and the impact of resolution on our results (Section~\ref{subsec:resolution-study}). We support the work in this section with comparisons to simulations of galaxies with more quiescent merger histories, showing that the observed differences are particularly evident in the merger scenarios. In Section~\ref{sec:discussion}, we suggest reasons why magnetic fields have been ineffectual in previous simulation work, discuss our results in the context of galaxy evolution as a whole, and discuss the main caveats to our results. Finally, in Section~\ref{sec:conclusions}, we summarise our main conclusions.

\section{Methodology}
\label{sec:methodology}

In order to observe the magnetic fields at their most effective, we have simulated a series of gas-rich major mergers. In particular, we have re-run the merger simulations first presented in \citet[]{sparre2016} with the inclusion of ideal MHD physics. These mergers were considered to be an ideal starting point as: 1) the progenitors had large, gas-rich discs, implying MHD analogues that would have strong, well-ordered magnetic fields; 2) at coalescence the gas densities were shown to reach high values, implying correspondingly strong amplification; and 3) the merger remnants in these simulations were able to rebuild their discs. The rebuilding process is naturally dependent on gas dynamics, allowing for further influence from the magnetic fields.

The set-up for these merger simulations is discussed in the following section. We discuss the set-up for the simulations of galaxies with more quiescent merger histories in Section~\ref{subsec:methods_isolated_galaxies}. In each case, the hydrodynamic and MHD simulations use the same underlying numerical implementation, such that a hydrodynamic simulation is equivalent to an MHD simulation with seed field set to zero.

\subsection{Initial conditions and parameters}

The initial conditions for the merger simulations are the same as those presented in \citet{sparre2016, sparre2017}. These were created by selecting four galaxies from the hydrodynamic cosmological simulation Illustris \citep{vogelsberger2014, vogelsberger2014b, genel2014} that had undergone a major merger between $z=1$ and $z=0.5$, were relaxed at $z=0$, and had a final stellar and halo mass close to that of the Milky Way. Zoom initial conditions were then created with a modified version of the \textsc{N-GenIC} code \citep{springel2015} for a periodic box of side 75 co-moving Mpc~$h^{-1}$, using the WMAP-9 \citep{hinshaw2013} cosmological parameters; i.e. the density parameters for matter, baryons and a cosmological constant: $\Omega_\text{m} = 0.2726$, $\Omega_\text{b} = 0.0456$, $\Omega_\Lambda = 0.7274$, respectively, and Hubble's constant $H_0 = 100$ $h$ km s$^{-1}$ Mpc$^{-1}$ with $h = 0.704$.

Dark matter particles were given a high-resolution within a roughly spherical region around the target galaxy, with a shell of standard resolution particles following, and lower-resolution particles filling the remaining volume. The highest dark matter mass resolution in the simulation was set to:

\begin{equation}
    m_{\text{DM}} = \left( \frac{1820}{2048 \times \text{`zoom factor'}}\right)^3 \times 6.299 \times 10^6 \; \text{M}_\odot,
    \label{eq:zoomfactor}
\end{equation}

where 1820/2048 is the ratio between the number of dark matter particles per box length in Illustris relative to our standard resolution\footnote{Standard resolution is therefore equivalent to setting `zoom factor' = 1.}, and $6.299 \times 10^6 \; \text{M}_\odot$ is the finest dark matter mass resolution in Illustris. Our simulations were run with zoom factors equal to 1, 2, and 3, which corresponds to dark matter mass resolutions that are 1.4, 11.4, and 38.5 times finer than in the original Illustris run. Equivalently, simulations with a zoom factor of 3 have $\sim1.8$ times finer mass resolution than the fiducial `Level 4' Auriga simulations \citep{grand2017}.

Following \citet{springel2005b} and \citet{price2007}, the softening length, $\epsilon_\text{DM}$, was chosen to be $\sim$$1/40$ of the initial average particle spacing:

\begin{equation}
    \epsilon_\text{DM} \approx \frac{L}{40 \times 2048 \times \text{`zoom factor'}},
    \label{eq:softening}
\end{equation}
where $L$ is the box length. The softening length is a co-moving length until $z=1$, at which point it is frozen in physical units, thereby maintaining the same resolution in the simulation for $z<1$. This helps to prevent unrealistic two-body interactions at early times, whilst still allowing small-scale structure to continue to form at late-times \citep[see, e.g.][]{power2003}. As gas cells vary strongly in density, their softening length is also scaled by the mean radius of the cell. Such cells have a minimum co-moving softening length of 30~$h^{-1}$~pc (frozen at $z=1$, as before) and a maximum physical softening length of 1.1 kpc. A full list of mass resolutions and softening lengths for each zoom factor is seen in Table~\ref{tab:sim_setup}.

\begin{table}
    \centering
    \caption{Zoom factor, finest dark matter mass resolution, finest baryon mass resolution, and softening length at $z=0$ for our highest, intermediate, and lowest-resolution runs, respectively.}
    \label{tab:sim_setup}
    \begin{tabular}{cccc}
        \hline \T
        Zoom factor & $m_{\text{DM}}$ [$\mathrm{M}_\odot$] & $m_{\text{b}}$ [$\mathrm{M}_\odot$] & $\epsilon_\text{DM}$ [kpc] \B \\
        \hline \T
        3 & $1.64 \times 10^5$ & $2.74 \times 10^4$ & 0.22 \\
        2 & $5.53 \times 10^5$ & $9.24 \times 10^4$  & 0.32 \\
        1 & $4.42 \times 10^6$ & $7.39 \times 10^5$  & 0.65 \B \\
        \hline
    \end{tabular}
\end{table}

\subsection{\textsc{Arepo}, Auriga, and MHD implementation} 
\label{subsec:set-up}

 The simulations were evolved from $z=127$ using the Auriga galaxy formation model \citep{grand2017} and the moving-mesh code, \textsc{arepo} \citep{springel2010, Pakmor2016I, weinberger2019}. \textsc{Arepo} uses a set of mesh-generating points to define a Voronoi tessellation, on which a second-order accurate, finite-volume Godunov scheme is formulated. Mesh-generating points may be moved arbitrarily and cells can be refined and de-refined such that they maintain a target mass resolution. In this manner, cells follow the flow of mass and thereby inherit the advantages of both Lagrangian and grid-based Eulerian codes. It has been shown that \textsc{arepo} is considerably more accurate than standard smoothed-particle hydrodynamic (SPH) methods when applied to a range of computational fluid dynamic problems \citep{sijacki2012} and produces the expected Kolmogorov turbulent cascade \citep{kolmogorov1941} for subsonic turbulence, unlike standard SPH models \citep{bauer2012}. The power spectrum of turbulence has a significant impact on its ability to excite the small-scale dynamo, making this result especially germane to our investigation.

The Auriga galaxy formation model is closely based on the models of \citet{vogelsberger2013} and \citet{marinacci2014} but contains important changes with respect to stellar feedback and the inclusion of the \citet{pakmor2013} MHD implementation (both summarised in the following text). The Auriga model has been shown to be able to produce Milky Way (MW)-like galaxies with appropriate stellar masses, sizes, rotation curves, star formation rates, and metallicities \citep{grand2017}, as well as finer details such as the correct structural parameters of bars \citep{calero2019} and the existence of chemically distinct thick and thin discs \citep{grand2018}. The models for star formation, stellar feedback, and active galactic nuclei (AGN) feedback are all physically well-motivated and parameters do not require retuning between resolution levels. Earlier work has shown that this is a non-trivial result \citep{scannapieco2012}. We summarise the main features of the model here, but encourage the reader to refer to \citet{grand2017} and references therein for a more comprehensive picture.

Gas may cool in Auriga via both atomic and metal-line cooling with self-shielding corrections accounted for \citep{vogelsberger2013}. A spatially uniform UV background field is included, which fully reionises hydrogen by $z \sim 6$ \citep{faucher2009}. The ISM is described using the \citet{springel2003} model, which treats star-forming gas as a two-phase medium governed by an effective equation of state. This model is derived from the assumption that, at the onset of thermal instability, processes below the resolution limit quickly lead to a pressure equilibrium forming between the hot and cold gas phases. It is not necessary to recalibrate this model when including magnetic fields. We show this explicitly in Appendix~\ref{appendix:ISM}.

Star particles are formed stochastically above a threshold density of $n_\text{SF}= 0.13\; \text{cm}^{-3}$ with a probability that scales with the local dynamical time. Each star particle represents a single stellar population, characterised by an age and metallicity, and assuming a \citet{chabrier2003} initial mass function. Stellar evolution is treated self-consistently, with mass loss and metal yields from supernovae SNII, SNIa, and asymptotic giant branch stars calculated at each time step and distributed to nearby gas cells using a top-hat kernel. The number of SNII events is set according to the number of stars formed in the mass range $8-100\;\text{M}_\odot$. This event is modelled by converting a star-forming gas cell into a wind particle and launching it in a randomly-chosen direction with a velocity proportional to the local one-dimensional dark matter velocity dispersion \citep{okamoto2010}. Wind particles interact only gravitationally until they reach a gas cell with $n < 0.05$ $n_\text{SF}$ or exceed the maximum travel time. The particle's energy is then deposited in the gas cell with the energy being split into equal parts thermal and kinetic. This results in smooth, regular winds, which become mostly bipolar at late times, as the wind takes the path of least resistance away from the galaxy. This is opposed to the bipolar wind model of \cite{marinacci2014}, in which wind particles are explicitly assigned an initial direction pointing away from the disc \citep[see, e.g.][for further details]{pillepich2018}.

Black holes are seeded with a mass of $10^5$ M$_\odot$ $h^{-1}$ in friends-of-friends (FoF) groups \citep{davis1985} with masses greater than $5 \times 10^{10}$ M$_\odot$ $h^{-1}$ at the position of the most dense gas cell. Black hole dynamics are governed by the \citet{springel2005a} model, with accretion described by an Eddington-limited Bondi-Hoyle-Lyttleton model and an additional term modelling radio accretion based on \citet{nulsen2000}. Feedback takes place through both radio and quasar modes, with thermal energy injected isotropically into neighbouring gas cells for the quasar mode, and bubbles of gas being gently heated at locations within the halo for the radio mode. The number of black hole neighbours are doubled with each increase in zoom factor in the standard way as a compromise between maintaining the total volume of neighbours and the increasing computational expense. In both quasar and radio mode, energy is injected continuously at a rate proportional to the accretion rate.

Magnetic fields are treated in the ideal MHD approximation \citep{pakmor2011, pakmor2013}, with the divergence constraint maintained through the use of a Powell 8-wave scheme \citep{powell1999}. In theory, the divergence constraint could also be preserved at machine precision using constrained transport schemes \citep{evans1988}, such as that implemented for \textsc{arepo} in \cite{mocz2014}. In practise, however, the Powell implementation performs sufficiently well, such that it is able to accurately replicate a series of MHD phenomena. These include: the linear phase of growth of the magneto-rotational instability \citep{balbus1991, pakmor2013}; the development of a small-scale dynamo in MW-like galaxies \citep{pakmor2014, pakmor2017}; similar field strengths and radial profiles to those observed in MW-like galaxies \citep{pakmor2017}; and Faraday rotation measure strengths that are broadly consistent to those observed for MW-like galaxies, both for the disc \citep{pakmor2018} and when compared with the current upper limits available for the circumgalactic medium \citep{pakmor2020}.

For each MHD simulation, we seed a homogeneous field of $10^{-14}$ co-moving Gauss, orientated along the $z$-direction, throughout the volume at $z=127$. This choice is essentially arbitrary as, for a broad range of values, all traces of the initial field strength and configuration are erased by an exponential dynamo in collapsed haloes \citep{pakmor2014}. Our choice of initial field strength has also been shown to produce magnetic fields that are dynamically irrelevant outside of collapsed haloes \citep{marinacci2016}. We note that during a star- or wind-forming event, the magnetic energy of the associated gas cell is removed, and is assumed to be locked-up in the subsequently formed stellar macro particle. Excluding this, magnetic fields are not explicitly included in our subgrid models.

\subsection{Galaxy tracking}
\label{subsec:galaxy_tracking}

In this work, we define haloes through the standard FoF approach and galaxies, or equivalently `subhaloes', using the \textsc{subfind} algorithm \citep{springel2001}. The distinction is useful as whilst FoF haloes may form tenuous bridges during galaxy interactions, causing them to be identified as a single structure, subhaloes are characterised by `self-boundness', meaning that structures remain essentially distinct until coalescence. The use of subhaloes hence allows us to consider the evolution of an individual galaxy until a very advanced stage of the merger. 

The primary and secondary progenitors of a merger are identified as the first and second most massive galaxies at $z=0.93$, as during this period both galaxies are relatively isolated (cf.\ \citealt{sparre2016}). In order to track the galaxies between snapshots, it is a case of identifying the galaxy that shows the most consistent trajectory to a previously identified one. In practise, this is the same as identifying the galaxy that contains the same black hole particle. Such a result is expected, as earlier work has shown that reliable merger trees can be constructed by tracking only the 10-20 most bound particles of each galaxy \citep{wetzel2009, rodriguez-gomez2015}. A short analysis of the deviations that occur is presented in Appendix~\ref{appendix:galaxy_tracking}.

\subsection{Simulations}
\label{subsec:sims}

\begin{table*}
\caption{Table of simulation parameters and merger remnant quantities at $z=0$ ($z=0.11$ for 1605-3). The columns show: 1) simulation run name; 2) physics included; 3) stellar mass ratio of main progenitors at $z=0.93$; 4) virial mass\color{blue}\protect\footnotemark[2]\color{black}; 5) virial radius; 6) bound stellar mass; 7) inferred stellar disc mass; 8) inferred stellar bulge mass;  9) disc-to-total stellar mass ratio\color{blue}\protect\footnotemark[3]\color{black}; 10) radial scale length; 11) bulge effective radius; 12) S\'{e}rsic index; 13) optical radius\color{blue}\protect\footnotemark[4]\color{black}; 14) gas-to-total baryonic mass fraction within the optical radius. We consider columns 3-6 and 14 to be global properties, whilst 7-13 are structural properties.}
\begin{tabular}{cccccccccccccc}
\hline
\textbf{Run}	&	\textbf{Physics}	&	$\frac{M_{*,1}}{M_{*,2}}$	&	$\frac{M_{200}}{10^{12}\;\mathrm{M}_\odot}$	&	$\frac{R_{200}}{\mathrm{kpc}}$	&	$\frac{M_{*}}{10^{10}\;\mathrm{M}_\odot}$	&	$\frac{M_\mathrm{d}}{10^{10}\;\mathrm{M}_\odot}$	&	$\frac{M_\mathrm{b}}{10^{10}\;\mathrm{M}_\odot}$	&	$D/T$		&	$\frac{R_\mathrm{d}}{\mathrm{kpc}}$	&	$\frac{R_\mathrm{eff}}{\mathrm{kpc}}$	&	$n$	&	$\frac{R_\mathrm{opt}}{\mathrm{kpc}}$	&	$f_\mathrm{gas}$	\\	\hline
\multicolumn{14}{c}{\T \text{Highest resolution} \B} \\
1330-3M	&	MHD	&	1.92	&	1.58	&	239.41	&	10.99	&	6.99	&	3.14	&	0.69	[0.41]	&	6.51	&	1.74	&	1.05	&	27.90	&	0.35	\\	

1330-3H	&	Hydro	&	2.00	&	1.52	&	236.18	&	11.07	&	6.60	&	2.12	&	0.76	[0.43]	&	7.00	&	1.70	&	0.69	&	15.43	&	0.16	\\	
1526-3M	&	MHD	&	1.08	&	1.75	&	247.74	&	5.72	&	3.18	&	1.88	&	0.63	[0.45]	&	4.25	&	1.72	&	0.92	&	18.23	&	0.17	\\	

1526-3H	&	Hydro	&	1.10	&	1.77	&	248.39	&	5.48	&	2.77	&	1.93	&	0.59	[0.35]	&	4.00	&	1.15	&	0.74	&	12.03	&	0.19	\\	
1349-3M	&	MHD	&	1.08	&	1.46	&	233.23	&	9.92	&	4.06	&	4.50	&	0.47	[0.43]	&	4.47	&	0.90	&	0.93	&	19.35	&	0.19	\\	

1349-3H	&	Hydro	&	1.11	&	1.45	&	232.51	&	9.43	&	4.27	&	2.82	&	0.61	[0.47]	&	4.28	&	0.95	&	0.72	&	13.14	&	0.09	\\	
1605-3M	&	MHD	&	1.29	&	1.07	&	203.60	&	7.98	&	3.26	&	3.11	&	0.51	[0.24]	&	1.72	&	0.85	&	0.66	&	11.56	&	0.19	\\	

1605-3H	&	Hydro	&	1.38	&	1.04	&	201.43	&	6.86	&	1.44	&	3.56	&	0.29	[0.11]	&	1.41	&	0.74	&	0.45	&	8.36	&	0.07	\\
\multicolumn{14}{c}{\T \text{Intermediate resolution} \B} \\
1330-2M	&	MHD	&	2.05	&	1.56	&	238.46	&	9.39	&	5.88	&	2.22	&	0.73	[0.44]	&	7.80	&	3.50	&	1.17	&	24.72	&	0.29	\\	

1330-2H	&	Hydro	&	2.06	&	1.59	&	239.89	&	8.73	&	5.54	&	1.88	&	0.75	[0.40]	&	6.35	&	2.68	&	1.17	&	23.06	&	0.31	\\ 
\multicolumn{14}{c}{\T \text{Lowest resolution} \B} \\
1330-1M	&	MHD	&	2.17	&	1.60	&	240.23	&	7.17	&	4.34	&	1.95	&	0.69	[0.39]	&	4.19	&	2.03	&	0.74	&	21.39	&	0.41	\\	

1330-1H	&	Hydro	&	2.20	&	1.54	&	237.20	&	6.88	&	4.10	&	1.56	&	0.72	[0.42]	&	6.46	&	2.70	&	1.05	&	21.61	&	0.31	\\	\hline
\end{tabular}
\label{tab:sim_data}
\end{table*}

In total, eight high, two intermediate, and two lower-resolution merger simulations were run. Each merger simulation is given a name with the format AAAA-BC, where AAAA is the four-digit FoF group number in Illustris for the halo containing the original galaxy at $z=0$, B is the `zoom factor' of the simulation, and C is the letter `M' or `H', denoting MHD or hydrodynamic physics, respectively.  The full list of simulations run is given in Table~\ref{tab:sim_data}. For the rest of the paper, references to a part of the run name implicitly refer to all simulations that contain this part -- e.g. a reference to 1330 refers to all simulations with this prefix.

Whilst we consider only one major merger event in each simulation, the trajectories and total number of participating galaxies vary. These differences have an impact on the production of the merger remnant, and so we briefly describe the interactions here. Roughly speaking, we may separate our merger scenarios into inspiralling (1330 and 1526) and head-on (1349 and 1605) major mergers. We proxy the beginning of the merger by the time of first periapsis. For 1330, this occurs at a lookback time of $\sim$5.4 Gyr ($z\approx0.54$), for 1526 it occurs at $\sim$6.8 Gyr ($z\approx0.77$), and for the 1605 and 1349 simulations, it takes place at $\sim$6.35 Gyr ($z\approx0.69)$. Every galaxy experiences an additional mix of minor mergers and fly-bys, with most of these accompanying the major merger. With this said, some relatively significant tidal interactions take place at approximate lookback times of 4 and 1 Gyr for 1330, 2.5 Gyr for 1526, 1 Gyr for 1349, and generally at late times for 1605. It should be noted too that the merger scenario in 1349 is particularly complex, with seven galaxies of significant mass existing within 100 kpc of the main galaxy at the time of the major merger. All of these galaxies, however, either coalesce with the main galaxy or leave its neighbourhood shortly following the major merger.

\subsection{Comparisons to more isolated galaxies}
\label{subsec:methods_isolated_galaxies}

In order to isolate the impact of mergers on our results, we compare our merger simulations to galaxies with more quiescent merger histories. For this purpose, we select four galaxies from the original Auriga \citep{grand2017} simulation suite (Au2, Au12, Au16, and Au23) and re-run these without magnetic fields. These galaxies have similar disc sizes and masses but all have significantly quieter merger histories than the galaxies in our simulations. This can be confirmed by observing their low accreted stellar mass fractions, $f_\text{acc}$, as given in Table 1 of \citet{grand2017}. To make sure that their growth is predominantly secular, Auriga galaxies are also selected such that they are farther than $9 \times R_\text{200}$ from any halo with a mass greater than 3 per cent of their own at $z=0$. The galaxy formation model used for these simulations is identical to that of our own. The dark matter mass resolution of these simulations is $3\times10^5\; \text{M}_\odot$, which is in between our zoom factor 2 and 3 runs. The softening lengths are therefore scaled accordingly (see Table 2 of \citealp{grand2017} for further details).

\section{Analysis}
\label{sec:analysis}

\subsection{Impact of MHD on global properties}
\label{subsec:MHD-global-properties}

\subsubsection{Halo and stellar mass}

In Table~\ref{tab:sim_data}, we provide a series of values that describe the merger remnant at the end of the simulation. These are given at $z=0.11$ for the 1605 simulations, as tidal disruption affects the structural properties of the remnant in the simulation at late times. For all other simulations, the data is given for $z=0$. From the table, it can be seen that the virial mass and virial radius of the merger remnant changes little, regardless of the resolution level or physics included in the simulation. Such a result is expected, as these statistics are dominated by the dark matter distribution, which feels no direct influence from magnetic fields and only a limited influence from any reorganisation of the baryonic matter. The similarity of these measures across all resolutions and physics models shows that the large-scale structure in our simulations is very well-converged. 

The small-scale baryonic structure, on the other hand, is slightly less well-converged between resolution levels. Here, the total stellar mass bound to the galaxy at the end of the simulation increases by a factor of roughly $\sim$25 per cent with each increase in zoom factor. A similar increase in stellar mass with resolution was also observed in \citet{grand2017}. In this case, it was noted that the excess mass mostly originated from stars born within the inner 5 kpc of the galaxy, where star formation is particularly susceptible to non-linear black hole accretion and related feedback loops \cite[cf.][]{marinacci2014}. Such processes will have been further affected by our treatment of black hole neighbours; as discussed in Section~\ref{subsec:set-up}, due to computational constraints the total volume of black hole neighbours decreases slightly with increased resolution. This will increase the initial energy density of the injected thermal energy.

Whilst there are clearly some small discrepancies as function of resolution, overall we consider the stellar mass values in our simulations to be sufficiently similar for the aims of this paper. This is especially so given the non-triviality of achieving convergence in cosmological simulations and the highly dynamic nature of the systems we have modelled. In particular, the stellar mass ratio of the main progenitors is sufficiently close that the results may be robustly compared with one another.

Considering the impact of including MHD physics, we see that this too has little effect on the final stellar mass bound to the galaxy. This is notable, as theoretically magnetic fields are able to provide pressure support to the gas, reducing the fraction above the threshold density, $n_\text{SF}$, thereby suppressing star formation. Some authors have also claimed evidence of a magnetically-driven wind arising in MHD galaxy simulations \citep[e.g.][]{steinwandel2019}, which would be able to remove cold, star-forming gas from the disc with the same effect. In contrast, we see no evidence of magnetic fields suppressing star formation in our simulations. Indeed, in most cases the final stellar mass is actually slightly higher in the MHD simulation compared to the hydrodynamic analogue.

\addtocounter{footnote}{+1}\footnotetext{Defined to be the mass inside a sphere in which the mean matter density is 200 times the critical density of the universe.}

\addtocounter{footnote}{+1}\footnotetext{Values are based on the circularity parameter defined in \cite{abadi2003}. The bracketed (unbracketed) values show the stellar mass ratio that kinematically belongs to the disc (doesn't belong to the bulge).} See Section~\ref{subsec:rot_support} for details.

\addtocounter{footnote}{+1}\footnotetext{Defined, as in \citet{grand2017}, as the radius at which the $B$-band surface brightness drops below $\mu_{B} = 25$ mag arcsec$^{-2}$. This region roughly encloses the disc.}

\begin{figure*}
\includegraphics[width=\textwidth]{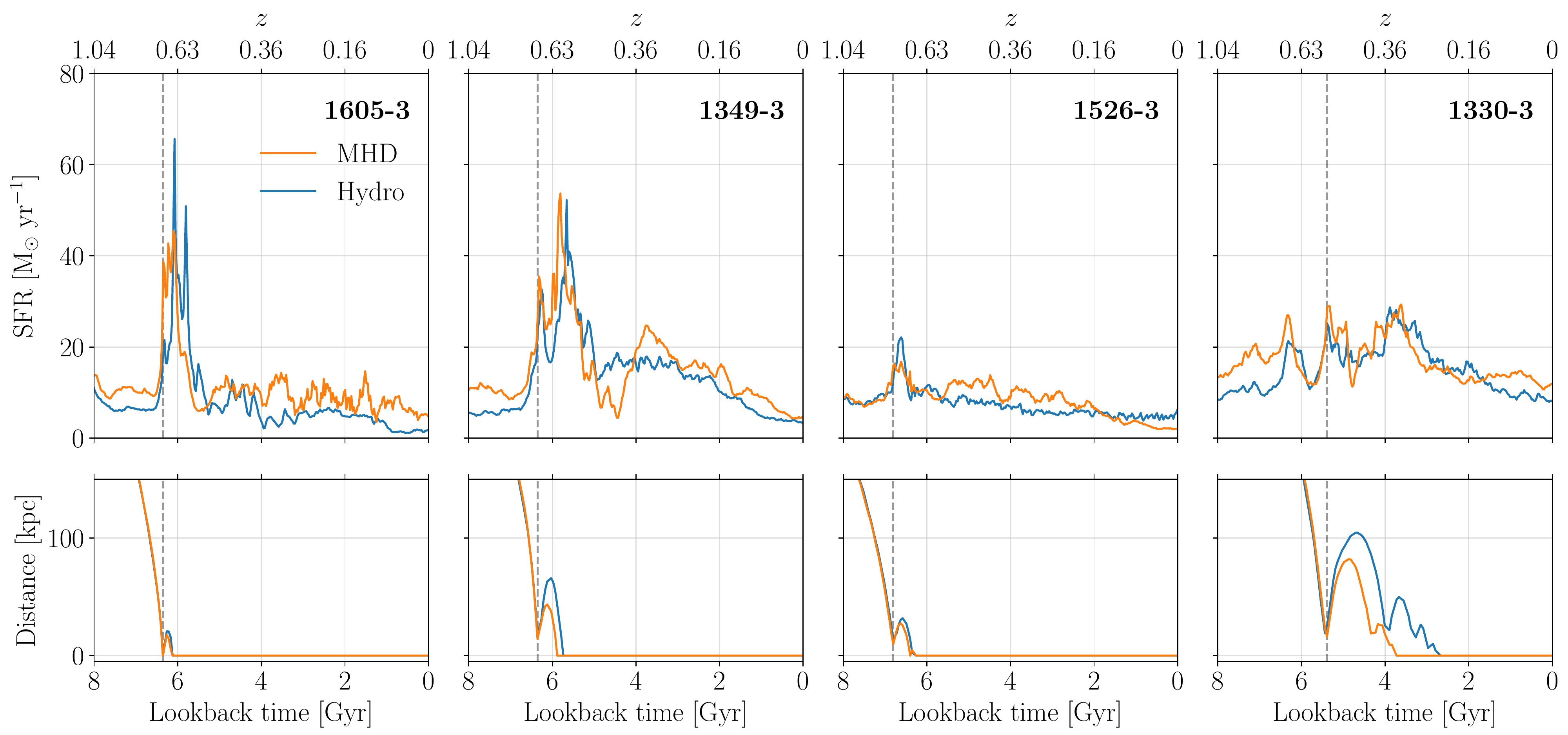}
    \caption{\textit{Top row:} star formation history for the main galaxy in each high-resolution simulation as a function of time. The dashed vertical line marks the time of first periapsis in the MHD simulations. \textit{Bottom row:} distance between the main progenitors as a function of time for the same simulations. The star formation history of the galaxy does not change significantly with the inclusion of MHD physics.}
    \label{fig:SFR_dist}
\end{figure*}

\subsubsection{Star formation history and orbit}

This picture is reinforced in Fig.~\ref{fig:SFR_dist}, where we show the star formation history for the main galaxy in each high-resolution simulation. To produce this, all star particles that were within $R_\text{opt}$\footnote{This radius encompasses virtually all stellar material in the galaxy. We have also conducted this analysis using a fixed radius of 30 kpc for each galaxy, which produces an essentially identical result.} (see penultimate column of Table~\ref{tab:sim_data}) of the centre of the main galaxy at $z=0$ were selected. The initial masses of these particles were then binned by their formation time, with bin widths set equal to 30 Myr. This width was chosen as it provides adequate time resolution, whilst preventing the data from becoming dominated by stochastic noise resulting from our probabilistic star formation model. This method is particularly advantageous as it is independent of the galaxy tracking process. A minor disadvantage, however, is that we exclude stars that have left the galaxy after formation. In practise, though, this has a negligible impact on the final result.

In each simulation, the merger causes a sudden rise in star formation, as existing gas is compressed and cold gas is brought into the galaxy. The timing of this rise correlates closely with the merging galaxy's periapsis, with the first approach generally causing the strongest burst. The merger scenarios presented in the two left-most panels are, broadly-speaking, the most energetic, being approximately head-on. Correspondingly, they show the most enhanced star formation. In contrast, the two right-most panels show inspiralling mergers. For these mergers, even the boosted star formation rate falls well short of the starburst threshold, as defined in \citet{sparre2017}. Significant star formation continues long after coalescence in every simulation, however, with none of the galaxies being quenched. This provides yet further evidence to support the argument that gas-rich mergers are not well-described by the `traditional' merger scenario, and instead preferentially produce a star-forming remnant.

Comparing simulations that included MHD physics to those that did not, we see that the magnitude, duration, and timing of the star formation peaks change very little. In fact, any discrepancies seen between the star formation histories can be more than adequately explained by the numerically stochastic nature of our star formation model, and by variations in the merger progression, as proxied by the distance between the two main progenitors.

Apart from indicating that magnetic fields have been ineffectual in suppressing star formation, the strong similarity of the star formation histories also provides further evidence that our simulations are numerically robust. This robustness is also seen in the almost identical evolution of the distance between the two main progenitors up until the first periapsis. After this time, the trajectories deviate a little. We note that the progenitors in the MHD simulations coalesce systematically faster than in the hydrodynamic analogues, and it is possible that this is an indication of more efficient transport of angular momentum in the MHD simulations. In general, however, the differences may also be explained by the non-linear N-body dynamics at play. In particular, the apparent extended orbit of the secondary galaxy in 1330-3H may be attributed to the late merger of its black hole and the subsequent continued identification of a distinct subhalo until this time.

\subsubsection{Amplifying the magnetic field}

\begin{figure*}
    \includegraphics[width=\textwidth]{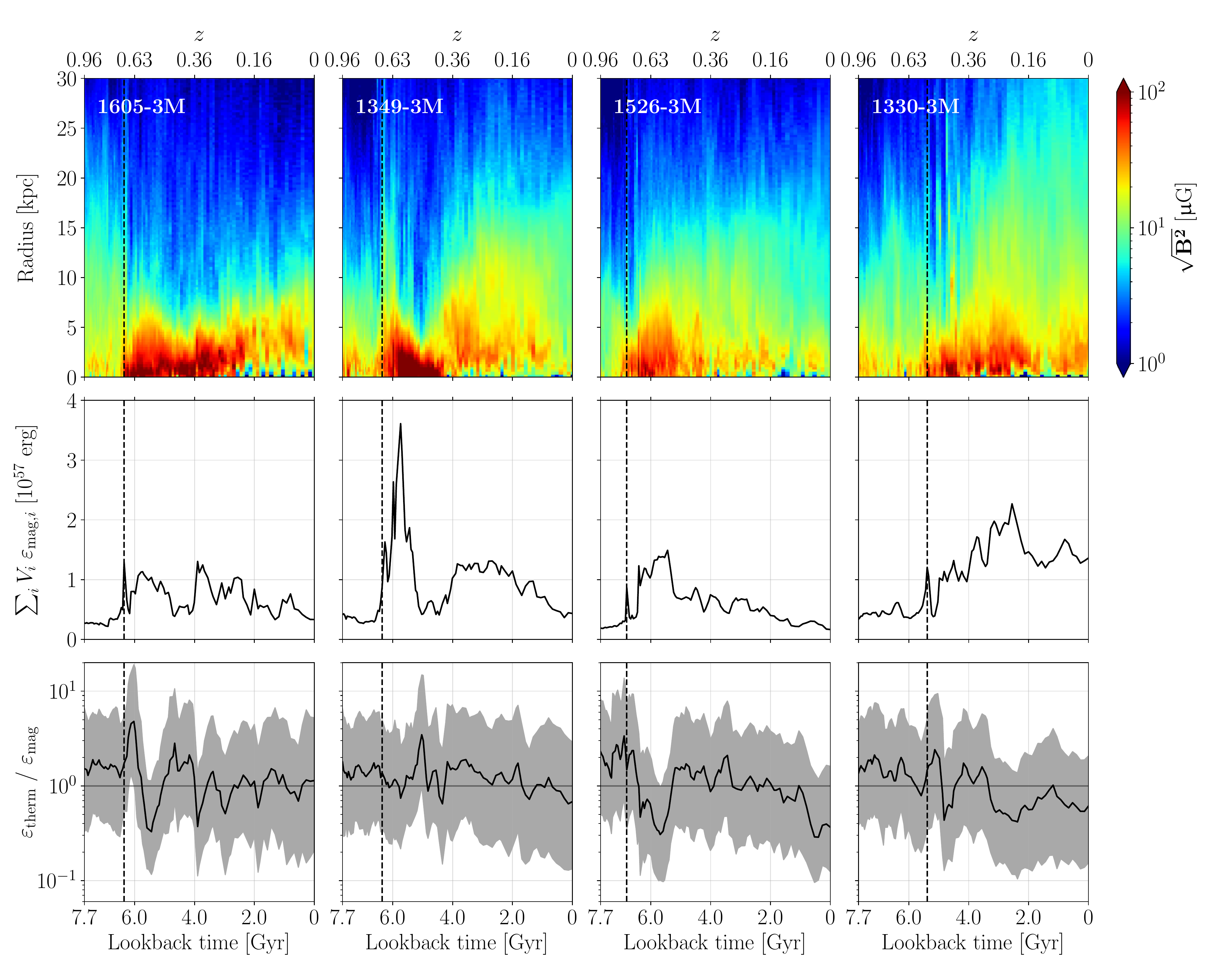}
    \caption{\textit{Top row:} radially-binned mean magnetic field strength in the galactic disc for each high-resolution simulation as a function of time. Bins have a radial extent of 0.25 kpc and a vertical extent of $\pm$1 kpc. The dashed vertical line marks the time of first periapsis. The merger in each simulation is able to substantially amplify the magnetic field in the inner 5 kpc by up to an order of magnitude, with effects visible for several Gyr afterwards. \textit{Middle row:} total magnetic energy in a disc with radial extent $R_\text{opt}$ (as given in Table~\ref{tab:sim_data}) and vertical extent $\pm$5 kpc. \textit{Bottom row:} distribution of the ratio of thermal-to-magnetic energy density for gas cells that lie within the same disc as above. The solid black line indicates the median, whilst the grey shaded region indicates the interquartile range. A horizontal line marks the point at which the magnetic energy density is equal to the thermal energy density. The merger generally results in a temporary, but significant, increase in the fraction of gas cells where the magnetic field is dominant.}
    \label{fig:BH_mag}
\end{figure*}

The broad similarity of the star formation histories produced by each physics model could easily be explained if the galactic magnetic field was not significantly amplified during the merger. However, this is not the case. In the top row of Fig.~\ref{fig:BH_mag}, we show the evolution of the radially-binned average magnetic field strength for each high-resolution simulation. To create this, volume-weighted means of the magnetic energy density are taken, using gas cells lying in annular rings of width 0.25 kpc and vertical extent $\pm1$ kpc; a region that covers the dense gas in the disc. The mean values are then converted back into an average field strength for the corresponding radius. It may be seen that immediately after first periapsis (indicated by the dashed black line) the magnetic field strength in the inner regions of the disc ($\lesssim5$ kpc) is strongly amplified by up to an order of magnitude. During this time the radial profile of the magnetic field strength continues to be well-fit by a double exponential, as observed in \citet{pakmor2017}. As expected, the strongest amplification of the magnetic field occurs for the most energetic mergers (1605-3M, 1349-3M). Indeed, for these galaxies a number of pixels over-saturate in Fig.~\ref{fig:BH_mag}. This is particularly the case for 1349-3M, where a few pixels reach over 150 $\upmu$G with mean field strengths reaching a maximum of $310$ $\upmu$G. Field strengths this high are unusual, but are not unheard of for starburst galaxies \citep[see, e.g.][]{lacki2013}.

The remaining field strengths in our simulations are in good agreement with those expected for gas-rich merging galaxies. In particular, the early stages of amplification seen for the inspiralling galaxies (1526-3M, 1330-3M) are consistent with strengths observed for the Antennae galaxies \citep{basu2017}, whilst the evolution until coalescence is consistent with that derived from nearby interacting galaxies by \citet{drzazga2011}. Our galaxies differ in their evolution post-merger, however, as whilst \citet{drzazga2011} predicts the galactic magnetic field to weaken significantly after coalescence, in our simulations the field remains highly amplified for at least 1.5 Gyr in each instance. Furthermore, when the field strength eventually does decrease, it returns to a strength that is at least as high as that which the galaxy had pre-merger. This difference in evolution is likely to be due to the different nature of the merger scenarios that we simulate; whilst the remnants in the Drzazga scenario are quenched, those in our own simulations are not. Instead, the remnants in our simulations maintain a significant percentage of their gas content, allowing them to also maintain the strength of their magnetic field until well after the initial merger-induced starburst has passed. Assuming our simulations reflect reality, this provides a new potential observable: an observation of an unusually high magnetic field in a galaxy that otherwise has a normal or low star formation rate could be an indication that the galaxy has undergone a gas-rich major merger in its recent past.

The decrease in magnetic field strength in the inner regions after the initial period of amplification is correlated with the rebuilding of the disc. This rebuilding is seen in Fig.~\ref{fig:BH_mag} through the increase in the field strength at larger radii. In some galaxies, at late times the field strength also reduces at these larger radii, as magnetic flux is locked up in newly-formed star particles and the consumed gas is not replenished. Such a process is seen particularly for 1526-3M from ${\sim}$2 Gyr onwards. Gas is also removed periodically from the inner regions due to AGN activity, causing the magnetic field strength to `flicker' at late times. This process may be undermining the amplification of the magnetic field generally, as field strengths do not generally recover to the same level afterwards. On the other hand, there are periods when the magnetic field strength increases at late times. For example, we see an enhancement of the magnetic field strength in 1330-3M by a factor of roughly two at $z \approx 0$, relative to its value at $\sim$1.5 Gyr. The galaxy in this simulation undergoes a series of minor tidal interactions at late times. It is not clear though whether the field amplification seen is a direct result of these interactions; whilst \citet{pakmor2017} do observe that minor mergers can cause such an effect, it is difficult to distinguish this particular enhancement from similar order fluctuations seen in the other simulations.

Although not shown explicitly here, at the time of the merger the field strength in the inner regions increases by more than that expected from pure adiabatic compression. This suggests that the amplification at this time is at least in part due to a small-scale dynamo. Evidence was shown in \citet{pakmor2017} that such a dynamo was active in the Auriga galaxies. In particular, it was shown that the kinetic energy power spectrum took on a characteristic \citet{kolmogorov1941} spectrum, producing in turn a magnetic energy power spectrum of the \citet{Kazantsev1968} type, as expected from small-scale dynamo theory. Naturally, the turbulent energy available in our simulations for such amplification should be even higher, owing to the strong solenoidal and compressive forcing during the tidal interactions We will explicitly study the existence of such a small-scale dynamo in Section~\ref{subsec:amplification_res}, showing that power spectra of the forms described here continue to be evident in our own simulations. 

In contrast to the inner regions, the magnetic field strength in the outskirts of the galaxies ($r \gtrsim$10 kpc) increases and decreases almost exactly with $\rho ^ {2/3}$. This implies that the field strength at these radii depends almost exclusively on flux conservation \citep{Kulsrud2005}, and that any dynamo that may exist is already saturated. This observation is further supported by the lack of amplification and field reorganisation seen after the merger at such radii, despite the passing of several Gyr. 

In the middle row of Fig.~\ref{fig:BH_mag}, we show the total magnetic energy in a volume with radial extent $R_\text{opt}$ and vertical extent $\pm5$ kpc. This region approximately bounds the disc and its immediate neighbourhood. It can be seen that the total magnetic energy generally follows the fluctuations seen in the radial evolution above. This is expected, as the amplification of the inner regions substantially contributes to the total energy in the volume. In each case, the magnetic energy spikes at the first periapsis as the energy of the merging galaxy is included in the calculation. The energy then increases more consistently shortly afterwards, as the associated turbulence and compression works to amplify the galactic magnetic field. This amplification is substantial, and can increase the total magnetic energy by up to an order of magnitude, as can be seen for 1349-3M. This period of heightened magnetic energy generally lasts for a shorter time than the corresponding inner amplification for most simulations, as magnetic energy decreases in the surrounding volume.

The period of initial amplification is generally followed by a second, longer period of increased magnetic energy. This is a result of the rebuilding of the gas disc in the remnant, and is once again particularly clear for simulation 1349-3M. For 1605-3M, this period is also a time of high activity from the central AGN. This results in a strongly non-linear evolution of the total magnetic energy, reflecting the subsequent fluctuations of the magnetic field strength in the inner $\lesssim5$ kpc. The two periods of increased magnetic energy are not very well separated in simulation 1330-3M. Here, the phases merge as the merger takes place over a sustained duration (see Fig.~\ref{fig:SFR_dist}). This means that the merging galaxy drives turbulence, and the resulting dynamo effect, over a period of several 100 Myr.

For most simulations, the total magnetic energy decreases towards the end of the simulation, returning to a roughly pre-merger level. At this time the turbulent driving from tidal interactions has long since stopped, and there is no longer a sufficient energy budget to maintain the amplified field strength. Once again, simulation 1330-3M does not quite follow this evolution, as it continues to be harassed at late times by satellite galaxies. Indeed, a particularly close encounter takes place at around $\sim1$ Gyr, coinciding with the peak seen in the total magnetic energy here. On top of this, this galaxy retains its gas content to a greater extent (as may be seen from its $f_\text{gas}$ value in Table~\ref{tab:sim_data}), allowing it to maintain its magnetic energy as well.

\begin{figure*}
    \centering
    \includegraphics[width=\textwidth]{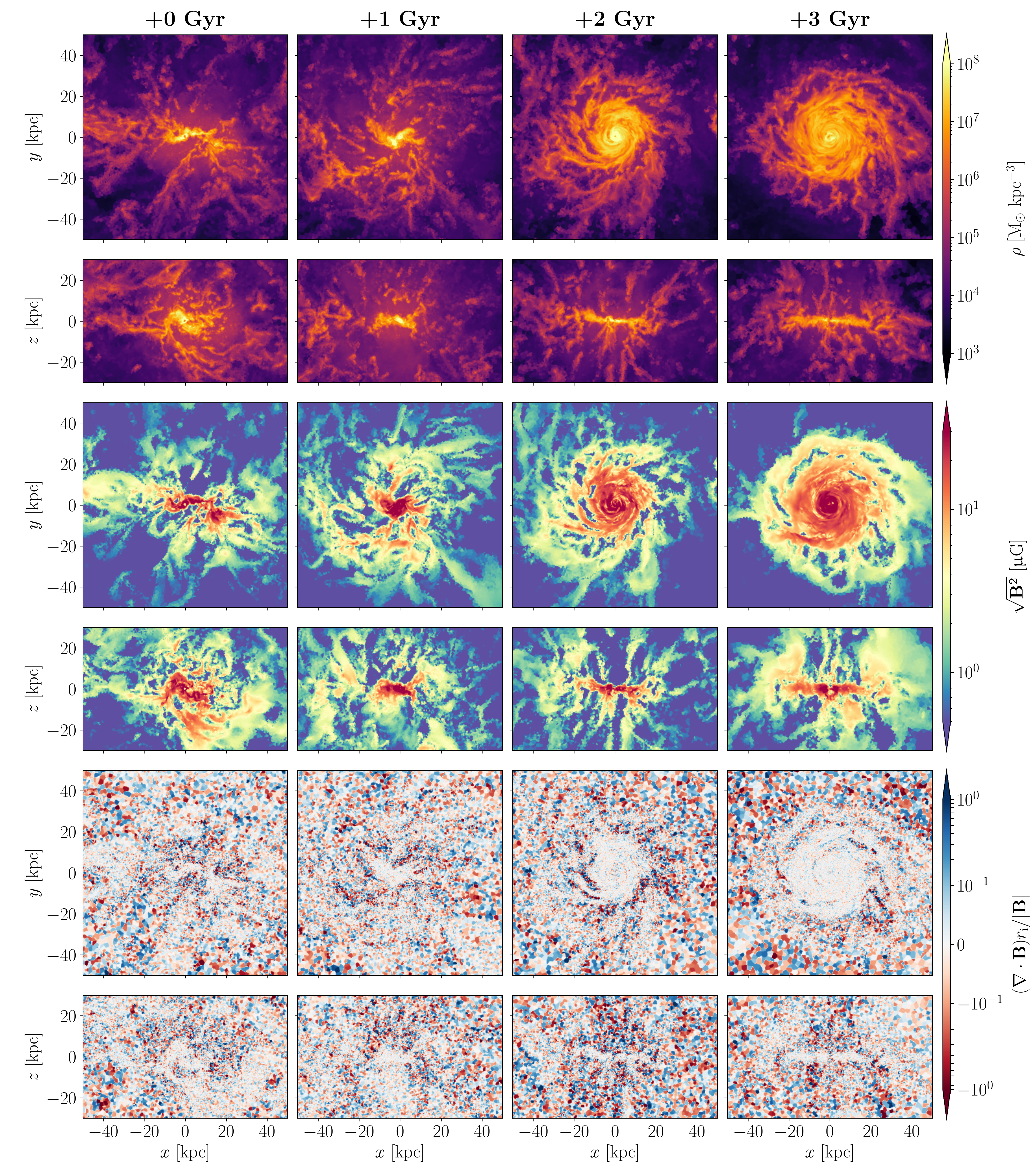}
    \caption{Face and edge-on slices through the main galaxy in simulation 1330-3M as it rebuilds post-merger. Timings are given from the first periapsis, showing a period when the galactic magnetic field is highly amplified. \textit{1st and 2nd row:} slices show the gas density in each cell. \textit{3rd and 4th row:} slices show the magnetic field strength in each cell. \textit{5th and 6th row:} slices show the relative divergence error in each cell. Divergence errors are highly localised and do not propagate between snapshots. Regions with higher gas density are better resolved due to our refinement scheme. These cells typically have the highest magnetic field strengths and the lowest relative divergence errors.}
    \label{fig:divergence_map}
\end{figure*}

The change in magnetic energy also changes the energetic balance of the system. In particular, it changes the ratio of thermal to magnetic pressure in the individual gas cells. This ratio varies strongly, both spatially and temporally, and is not well-captured by a radial average. Instead, in the bottom row of Fig.~\ref{fig:BH_mag}, we consider the distribution of this ratio for gas within the same volume as above. The distribution can be strongly skewed by extreme values, and so we show the median and interquartile range, rather than the mean. Larger values of this statistic imply that the gas dynamics are more affected by the thermal component, whilst smaller values imply that the magnetic fields are more influential. 

It can be seen that at all times, the gas cells cover a broad range of values, indicating that there are regions throughout the galaxy where either the thermal or magnetic pressure is dominant. In each case, however, the distribution is biased towards thermal pressure at early times, indicating that magnetic fields are, on the whole, subdominant at this time. The arrival of the secondary progenitor results in gas in the galactic neighbourhood being compressed and the production of a large amount of turbulence. This initially increases the fraction of thermal pressure, before amplification of the magnetic field swings the distribution the other way. This development takes place within a few 100 Myr, consistent with the time-scales required for a small-scale dynamo to amplify the field \citep{arshakian2009}. This evolution is less clear in 1349-3M, which may be a result of its more complex merger scenario (see Section~\ref{subsec:sims}). After the initial period of amplification, the evolution of the distribution is highly non-linear, depending strongly on the stability of the magnetic field. As noted previously, this is seen particularly in 1605-3M, where AGN outbursts lead to oscillations in the balance of the thermal to magnetic pressure distribution. Generally, the fraction of magnetic pressure relative to thermal pressure has increased by the end of the simulation.

In Fig.~\ref{fig:divergence_map}, we show how the disc regrows after disruption. In the top and middle two rows, respectively, we show the gas density and magnetic field strength as seen in slices through the main galaxy in simulation 1330-3M. We present four times showing the development of the galaxy, starting with its state at first periapsis, with snapshots thereafter showing progressive states at increasing 1 Gyr increments. During this period, the magnetic field is at its most amplified. The angular momentum of the merging galaxy in this simulation is particularly well-aligned with the main galaxy, and consequently the disc grows quickly within the time-frame shown. The development seen is, however, qualitatively similar for all our merger simulations.

It is clear from both the gas and magnetic field strength distributions seen in the first column that the galaxy is substantially disrupted by the approach of the merging galaxy. Such disruption already generates high field strengths through compression, before a dynamo has had time to saturate. The impact of the tidal interaction scatters the gas content of the galaxy, producing filaments of high density gas well below the disc plane. In the following snapshot, this gas has begun to accrete onto the galaxy, starting the process of disc-rebuilding and increasing the magnetic field strength at the core. Already at this stage, the magnetic field has begun to take on a relatively axisymmetric profile, which only becomes smoother with time. This justifies our choice of showing radial profiles in the top row of Fig.~\ref{fig:BH_mag}. The magnetic field is generally most dominant when it is strongest, and so the dynamics at the very centre of the disc will be particularly affected. This has further ramifications, as the evolution of the galaxy as a whole is sensitive to the behaviour of the central AGN, fed by the gas in this region. The exact impact of the magnetic fields on both the central gas distribution and black hole accretion rate will be explored in an upcoming paper.

\subsubsection{Stability of MHD implementation}

\begin{figure}
    \centering
    \includegraphics[width=\columnwidth]{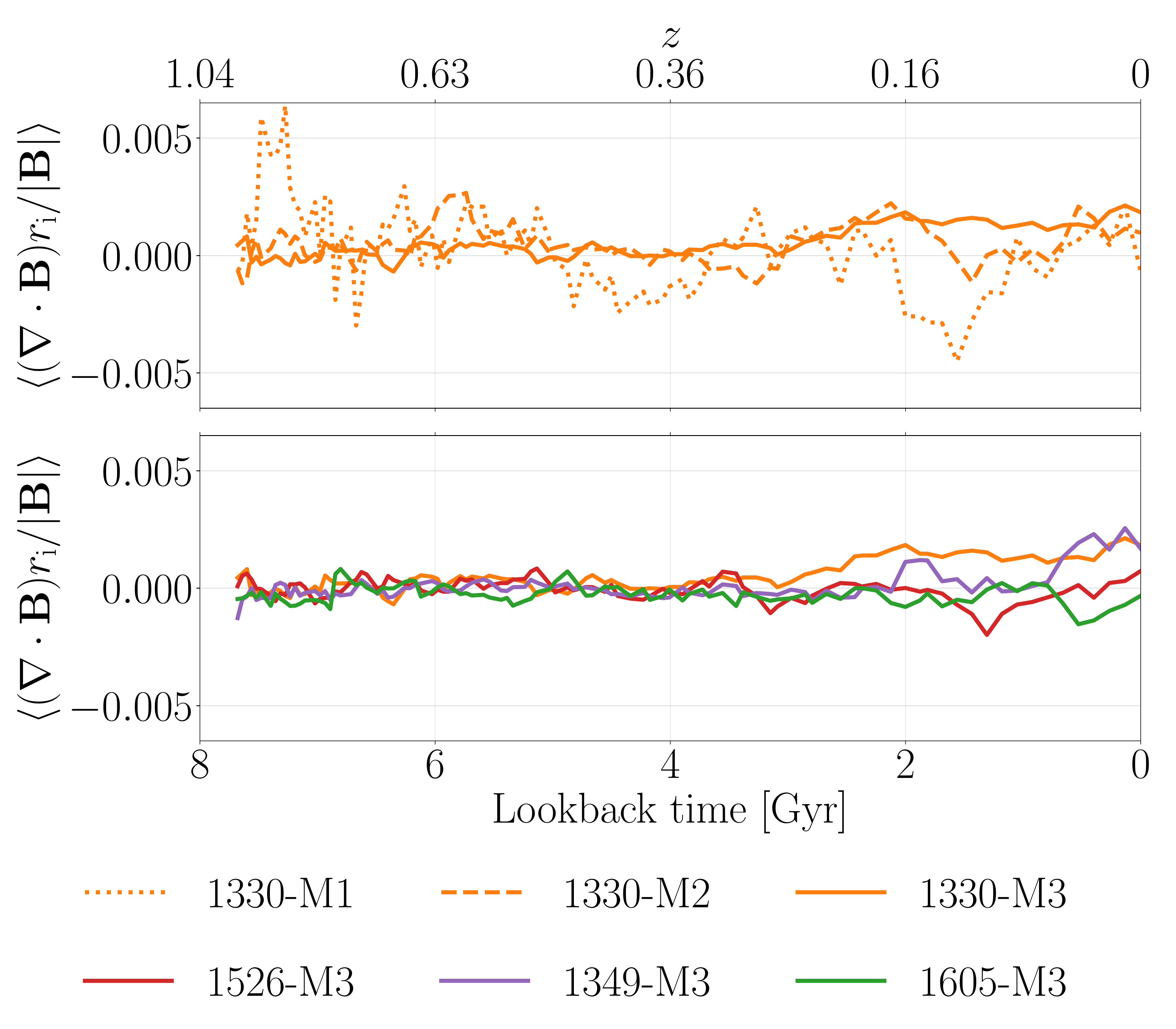}
    \caption{\textit{Top panel:} the mean of the relative divergence error, as a function of time, for simulations run with different resolution. \textit{Bottom panel:} as above, but for the highest-resolution simulations only. The mean is calculated for all gas cells that lie within 100 kpc of the main galaxy. It stays well below 1 per cent in each simulation and decreases with increased simulation resolution.}
    \label{fig:divergence}
\end{figure}

Given the strong amplification of the magnetic field seen in our simulations and the ability of the field to subsequently play an important dynamical role, it is prudent to consider the evolution and impact of divergence errors. Whilst the continuum equations of MHD preserve the $\bs{\nabla} \bs{\cdot} \bs{B} = 0$ condition perfectly given an initial divergence-free field, this is not the case for the discretised versions of the equations used in our simulations. Worse still, partial differential equation solvers are generally unstable to the production of magnetic monopoles; once produced, these have a tendency to become rapidly larger in any non-trivial MHD flow, rendering simulation results unphysical \citep{pakmor2011}. As discussed in Section~\ref{subsec:set-up}, divergence errors in our simulations are controlled using a Powell 8-wave scheme. Whilst it has been shown in previous work that this scheme is able to deal with divergence errors robustly \citep{pakmor2013}, it is sensible to analyse its performance in our simulations as well.

In the bottom third of Fig.~\ref{fig:divergence_map}, we show the distribution of the relative divergence errors as seen in a slices through the main galaxy in simulation 1330-3M. As with the panels above, this picture is qualitatively the same for all simulations. Whilst some large individual errors may be observed, they are highly localised in space. Furthermore, as observed in \citet{pakmor2013}, the sign of the divergence error is seen to alternate between neighbours when its magnitude becomes large. This alternation means that fluctuations generally cancel when considered over larger scales. Particularly large divergence errors generally stem from larger, under-resolved gradients in the local magnetic field. These, in turn, are often a due to larger cell sizes. Such cells are low density, as our refinement scheme keeps gas cells within a target mass. In contrast, the high density regions are very well resolved, and have subsequently lower relative divergence errors. This is true for a range of magnetic field strengths and means that the galactic disc, where the most substantial amplification occurs, is particularly robust to such errors. Perhaps even more importantly, the divergence errors are not seen to propagate in time. Instead, the distribution of errors in each snapshot is broadly independent of its previous distribution.

In Fig.~\ref{fig:divergence}, we show the mean of the relative divergence error in each simulation as a function of time. To calculate this, we select gas cells that lie within a radius of 100 kpc of the main galaxy at each time step, as this is well within the region of high-resolution in our simulations and also covers a volume that can affect the immediate development of the galaxy. We show the mean of the relative divergence error as we are interested in the stability of the system, rather than any particular peak values. Indeed, the Powell scheme implemented in \citet{pakmor2013} often produces higher average divergence errors than the Dedner scheme \citep{dedner2002} it replaced. The advantage of the Powell scheme, however, lies in its more effective control of such errors, which makes it more appropriate for cosmological simulations and highly dynamical systems. It can be seen in Fig.~\ref{fig:divergence} that this stability remains in our merger simulations. Indeed, the average divergence error decreases with increased resolution, as previously observed in \citet{pakmor2013}. This is in contrast to the magnetic field amplification, which increases with increased resolution (see Section~\ref{subsec:resolution-study}). We note on top of this that there are no signs of instability developing at the time of amplification (compare to Fig.~\ref{fig:BH_mag} and Fig.~\ref{fig:mag_zoom}), indicating that these are not the source of the amplification. We conclude from this that our results are robust to our MHD implementation.

\subsubsection{Bound gas and stellar mass evolution}

\begin{figure*}
    \includegraphics[width=\textwidth]{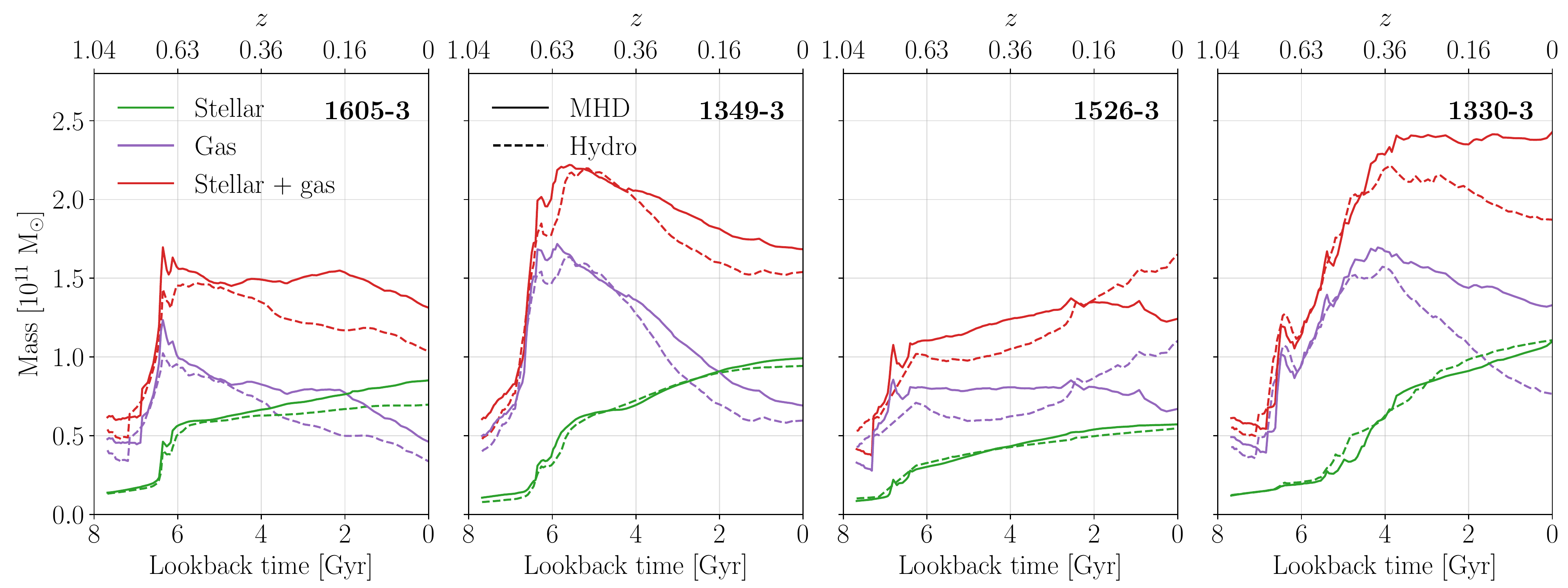}
    \caption{Total stellar and gas mass bound to the main galaxy in each simulation as a function of time. Whilst the stellar mass remains similar across physics models, the bound gas mass at $z=0$ is generally higher in MHD simulations. Indeed, in some cases, almost all gas lost can be accounted for by a comparable increase in stellar mass.}
    \label{fig:bound_gas_stellar}
\end{figure*}

Whilst the star formation history is generally very similar regardless of physics models used, the amplified magnetic fields are nevertheless able to significantly affect the gas dynamics. In Fig.~\ref{fig:bound_gas_stellar} we show the total gas and stellar mass bound to the main galaxy in each high-resolution simulation as a function of time, as well as the sum of these quantities. For each simulation, the merger results in a sharp increase in the total gas mass bound to the system, followed shortly thereafter by an increase in the bound stellar mass. The timing here is dependent on the rate at which the progenitors coalesce, as well as on the star formation history of the main galaxy. Both the gas and stellar mass evolutions exhibit a few localised peaks before full coalescence, as mass is reallocated between the merging galaxies by \textsc{subfind} \citep[cf.][]{rodriguez-gomez2015}. As the gas component is relatively diffuse, it is particularly sensitive to this reallocation process. This effect is seen for all simulations, but is most clear for the simulations where the merger took the longest. At the time of coalescence, the gas mass bound to the main galaxy is generally higher for the MHD simulations than for the hydrodynamic simulations. This supports the idea that the accelerated coalescence seen in Fig.~\ref{fig:SFR_dist} may be a result of more effective gas transport in MHD simulations at the later stages of the merger.

Assuming the galaxies exist in relative isolation post-merger, the gradient of the `stellar + gas' line provides us with information on how effective feedback is in ejecting gas from the galaxy: if gas is only being converted into stellar mass, this line will stay constant; if feedback is efficient, then this line will show a negative gradient, as gas is unbound from the system. Of course, the nature of cosmological simulations is that the galaxies do not experience complete isolation. We therefore expect to see some gas accretion after the merger; an effect that would be absent in idealised simulations. The accretion in our simulations takes place both through cosmological filaments and through further galaxy interactions, as mentioned in Section~\ref{subsec:sims}. This effect is seen most obviously where the sum of bound stellar and gas mass continues to increase after the merger, as in 1526-3. The hydrodynamic simulation in this case shows a particularly strong increase in gas mass at late times as gas is stripped off a passing galaxy. The evolution of the stellar and gas mass is also affected by the allocation of mass by \textsc{subfind} during such interactions. In particular, this effect results in the flattening out of the bound mass evolution seen for 1349-3H around 1 Gyr and the loss of bound mass starting at 2 Gyr for 1605-3M.

Despite such deviations, it is still apparent that for virtually every merger, feedback removes gas more effectively in the hydrodynamic simulation than in the MHD analogue. Indeed, for many of the MHD simulations, feedback is highly ineffective at unbinding the gas from the galaxy; frequently, the loss of gas mass may be accounted for almost entirely by the corresponding increase in stellar mass. This evolution is especially clear for simulation 1330-3M, but may also be seen in 1526-3M and 1605-3M. This provides further evidence that the winds in our simulations are not magnetically-launched. It is especially notable that we do not see magnetically-driven winds given the strength of the field reached in the disc post-merger. This result conflicts with the idealised smoothed-particle MHD simulations performed by \citet{steinwandel2019}, which did show such a wind, despite having a similar dark matter mass resolution. Whilst some differences will be due to the cosmological nature of our simulations, we suspect that the choice of numerical treatment also plays a significant role.

Finally, it is evident from Fig.~\ref{fig:bound_gas_stellar} that the differences in star formation seen in Fig.~\ref{fig:SFR_dist} do not accumulate significantly over time. Whilst the total bound gas mass diverges between physics models, the total bound stellar mass stays similar. The result is that the ratio of gas to total baryonic mass in the merger remnant (a measure of how gas-rich the remnant is) can differ substantially between the the two physics models by the end of the simulations. This fraction is given in Table~\ref{tab:sim_data} for a reduced radius, showing that the loss of gas does not only affect the outer regions of the circumgalactic medium. The mechanism behind the more effective unbinding of the gas in hydrodynamic simulations can begin to be understood by considering the structural properties of the merger remnants.

\subsection{Impact of MHD on structural properties}
\label{subsec:MHD-disc-sizes}

\subsubsection{Magnetic field and altered gas morphology}

\begin{figure*}
    \includegraphics[width=\textwidth]{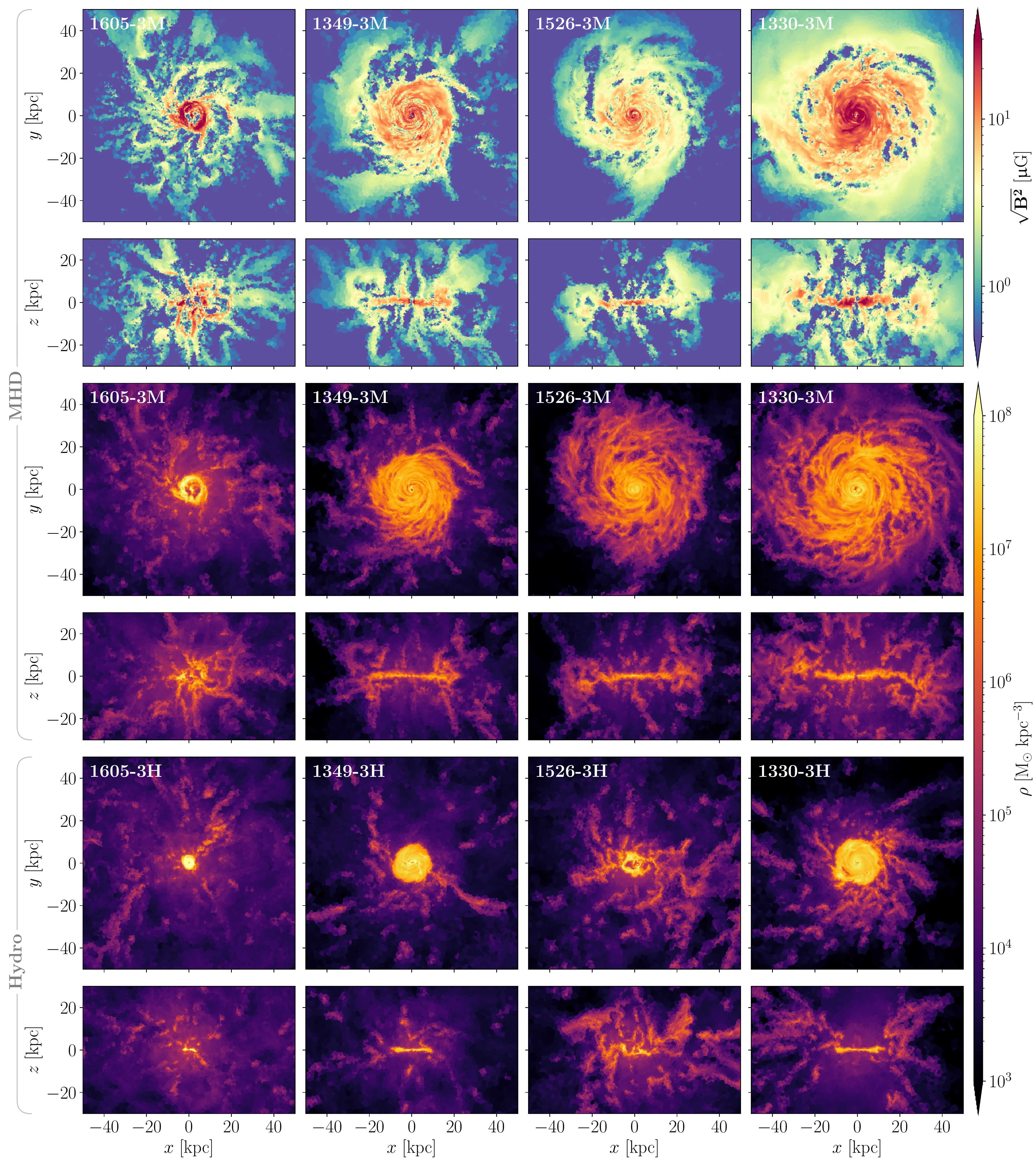}
    \caption{Face and edge-on slices through the merger remnant for each high-resolution simulation, as seen at $z=0$ ($z=0.11$ for 1605-3).  \textit{1st and 2nd row:} slices show the magnetic field strength in each gas cell. The magnetic field is broadly axisymmetric, but still shows significant amounts of small-scale structure. This roughly mirrors the corresponding gas distribution. \textit{3rd and 4th row:} slices show gas density for the MHD simulations. The gas discs have a flocculent structure, and show a shallow radial gradient. \textit{5th and 6th row:} slices show gas density for the hydrodynamic simulations. The gas discs are systematically smaller and thinner. They also exhibit a flatter density profile, which cuts off abruptly at the disc edge. The clearance of gas above the disc implies more effective stellar feedback is in action in these simulations. Simulations 1605-3M and 1526-3H have a more unusual morphology, owing to the impact of strong AGN feedback.}
    \label{fig:gas}
\end{figure*}

In Fig.~\ref{fig:gas}, we show slices taken face and edge-on through the main galaxy in each high-resolution simulation. These are shown at $z=0.11$ for the 1605 simulations (see Section~\ref{subsec:MHD-global-properties}), and at $z=0$ for all others. In the top two rows, the slices show the magnetic field strength in each gas cell, whilst the bottom four rows show the gas density for MHD and hydrodynamic simulations, respectively. As in Fig.~\ref{fig:divergence_map}, the magnetic field profiles are observed to be generally axisymmetric on large scales. With this said, it is clear that there is a great deal of small-scale structure to be found as well. In general, this structure mirrors the flocculent nature of the underlying gas disc, with regions of dense gas correlating to regions of high magnetic field strength. This is also true for regions above and below the disc, where dense clumps are still seen to be reasonably strongly magnetised, with field strengths on the order of a few $\upmu$G.

Whilst increased density is correlated with stronger magnetic fields, the reverse effect may be seen at the centre of the galaxies. Here, quasar feedback temporarily removes gas, weakening the field. This effect is particularly clearly seen for simulation 1605-3M in Fig.~\ref{fig:gas}, where the AGN has pushed gas out of the inner $\lesssim3$ kpc, helping to produce a ring shape morphology in the face-on view, and strongly distorting the disc in the edge-on view. For the other MHD simulations, the gas discs show shallow radial and vertical gradients. The gas disc for these galaxies extends in radius well beyond the stellar disc (c.f. Fig.~\ref{fig:mocks}). At these radii, a network of filamentary gas structures with densities below $10^7\;\text{M}_\odot\; \text{kpc}^{-3}$ can be seen. These structures show gas joining the galaxy in the plane of the disc, fuelling its radial growth. The clumps of gas seen above and below the disc are typical of a fountain flow in action. Such fountain flows have also been found to be crucial for successfully growing large galactic discs in previous work \citep{grand2019}.

The merger remnants formed in the hydrodynamic simulations are systematically smaller than those formed in the MHD simulations. Furthermore, they do not display the same level of complex small-scale structure as their MHD analogues. Rather, the gas density stays high throughout the disc, with a sudden cut-off seen at the disc edge, where the density drops by some three orders of magnitude. The amount of flocculent structure seen in the face-on view beyond the stellar disc is greatly reduced too, implying that these galaxies are not receiving the high angular momentum gas they need to grow in size. This understanding will be explicitly confirmed in an upcoming work.

As well as the reduced radius, the height of the disc in the simulations without magnetic fields has also been affected, with gas discs becoming razor-thin. The lack of disruption seen in the centre of the disc implies that this is not due to AGN feedback. Indeed, the sudden cut-off in gas density in the vertical direction suggests that the disc height has been affected by stellar feedback, with wind particles coupling to the low-density gas in the region, quickly removing it. This view is supported by the reduced gas density seen above and below the disc -- taken to its extreme in 1330-3H, where gas is cleared out of a conical-shaped region -- and by the lack of disruption to the disc midplane, where the gas density is too high for our simulated wind particles to effectively couple (see Section~\ref{subsec:set-up} for details). It is also noticeable that the gas is more effectively cleared above the disc for galaxies that have a higher stellar mass (see Table~\ref{tab:sim_data}). The increased effectiveness of the stellar winds in this case is logical, as more star formation takes place in a similar volume, leading to higher wind energy densities. Such energy densities help to explain the substantial unbinding of gas seen for hydrodynamic simulations in Fig.~\ref{fig:bound_gas_stellar}. As well as unbinding the gas, the stellar winds also help to maintain the reduced disc sizes; by clearing the gas above and below the disc, they disrupt small-scale fountain flows, further suppressing growth.

The only hydrodynamic simulation that does not fit the pattern is 1526-3H. Here, the merger remnant had a higher bound gas mass than its MHD analogue at the simulation end (see Fig.~\ref{fig:bound_gas_stellar}). However, this remnant also shows a clearly different gas morphology, displaying similar disruption to that seen for 1605-3M. This disruption indicates recent AGN feedback has taken place in this galaxy as well, and generally indicates a different evolutionary path taken compared to the other hydrodynamic simulations.

\subsubsection{Altered stellar morphology}

\begin{figure*}
    \includegraphics[width=\textwidth]{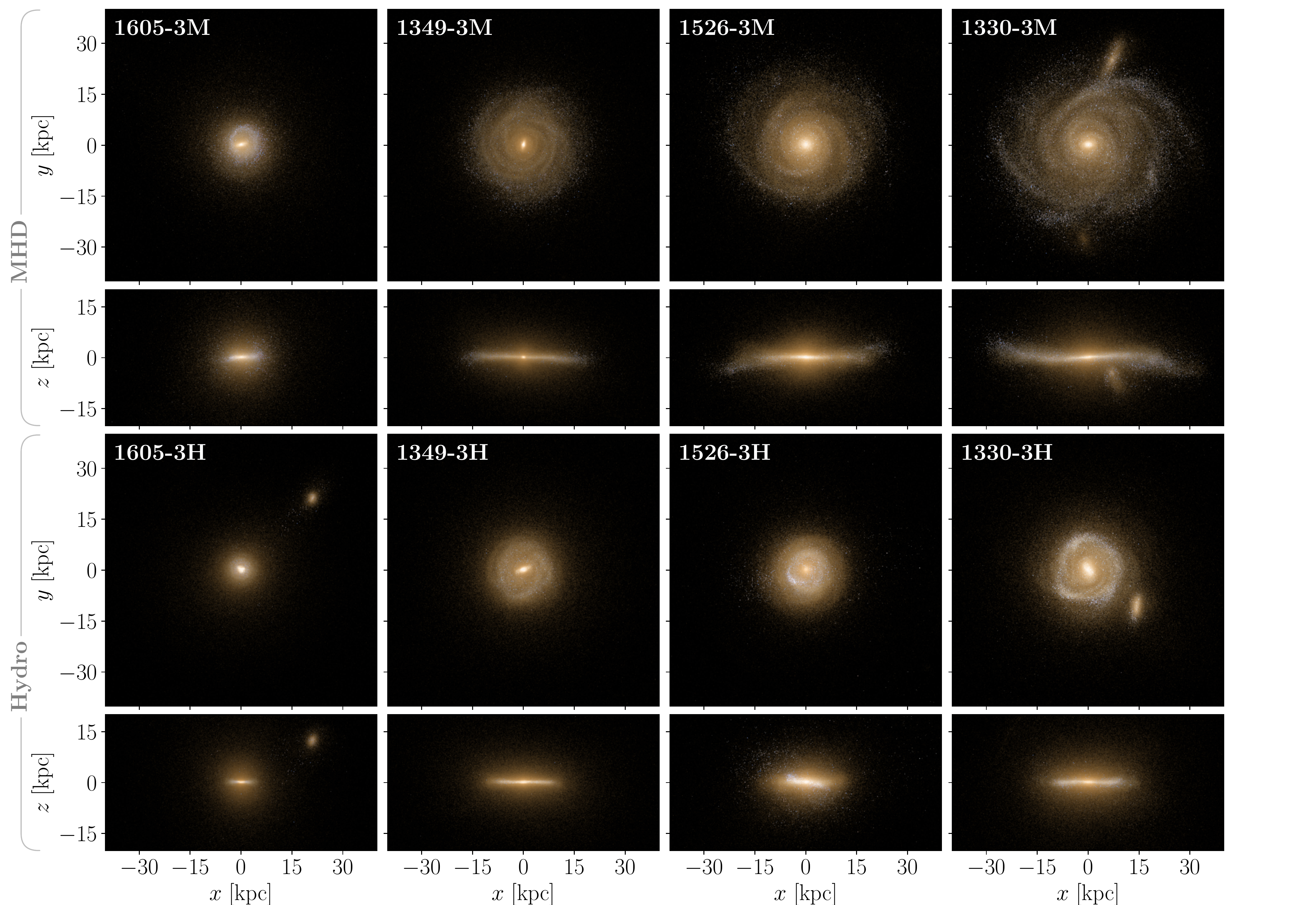}
    \caption{\textit{1st and 2nd row:} mock SDSS \textit{gri} composite images showing the merger remnants for all high-resolution MHD simulations. Remnants are seen face and edge-on at $z=0$ ($z=0.11$ for 1605-3). \textit{3rd and 4th row:} as above, but for the hydrodynamic simulations. The morphology of the merger remnant is once again systematically different between MHD and hydrodynamic runs. Whilst MHD simulations generally produce MW-like galaxies that show extended spiral-structure, hydrodynamic simulations produce a more compact disc with unusual stellar bar and ring features.}
    \label{fig:mocks}
\end{figure*}

The different gas morphologies naturally lead to different stellar morphologies. To analyse these, we produce a series of mock observational images, using the estimated photometric properties of the star particles. These properties are calculated using stellar population synthesis models based on data given in \citet{bruzual2003} and are provided in the form of mock SDSS broad band luminosities. Following \citet{vogelsberger2014}, we map the $g$, $r$, and $i$-band luminosities to the red, green, and blue channels of an RGB image, binning each channel to create a projected image. The resultant values are then scaled according to algorithms presented in \citet{lupton2004}.  A transparency factor is also set proportional to the maximum binned $g$-band luminosity. The final image does not include effects such as dust attenuation, and is therefore not a true observational mock, but it nevertheless provides much useful information. In particular, it allows us to easily identify prominent morphological features in the remnants. In Fig.~\ref{fig:mocks}, we present face and edge-on mock images of the merger remnant in all high-resolution simulations, as created in this manner. Once again the remnants are seen at $z=0$, except for the 1605 simulations, which are shown at $z=0.11$. For each image, we use the data from all star particles that exist within a depth of $\pm40$ kpc.

As expected, many of the features that were visible in the gas morphology are seen here as well. In particular, the stellar discs produced in MHD simulations are systematically larger than their hydrodynamic counterparts. Naturally, this difference is largest for the largest remnants. The discs in MHD simulations are also thicker and less sharply-defined, following the distribution of the gas. Indeed, the gas distribution is generally well-reflected in the stellar light distribution; for example, the flocculent gas structure observed for the MHD simulations in Fig.~\ref{fig:gas} is seen to support a significant amount of spiral structure here. In contrast, remnants from the hydrodynamic simulations, which had much less small-scale gas structure, show no evidence of spiral arms. Instead, they display distinctive bar and ring morphologies, more typical of barred lenticular galaxies. Morphologies of this kind are not unheard of, but are also certainly not usual for MW-size galaxies. Where they do exist, the rings are often theorised to be a result of resonant forces channelling the gas. This, in turn, is sometimes interpreted as evidence that the galaxy has undergone a mostly secular evolution \citep[e.g.][]{buta2004}. Our simulations show, however, that this must not necessarily be the case.

Once again, simulations 1605-3M and 1526-3H do not quite fit the pattern seen in the other simulations. The chaotic gas dynamics shown in Fig.~\ref{fig:gas} are here reflected by the diffuse interior stellar rings and puffed-up discs. Mergers and tidal interactions have been shown to be able to puff up stellar discs in previous work \citep[e.g.][]{welker2017}, but this is unlikely to be the case here. In particular, we note that this morphology is not seen for the other simulations, even when the remnant has experienced an interaction recently, such as in 1605 and the 1330 simulations, where interloping satellites can be seen in the mock images. Instead, this particular morphology is likely to be a result of gas being lifted above the disc midplane by AGN outbursts. Such outbursts also likely explain the diffuse nature of the ring; star formation is triggered at the edges of the outflow region where gas piles up. However, these outbursts are typically irregularly-shaped and are not consistent over time, meaning that star formation is not reinforced at the same radius as it is for the other ring galaxies. The merger remnant in 1526-3H has a particularly large scale height, which is likely to be a result of the orbit of its central black hole. This is not well-tied to the galactic centre, as in the other simulations, meaning that the AGN feedback is consequently not well-localised. We explore this issue further in Appendix~\ref{appendix:galaxy_tracking}.

\subsubsection{Rotational support and surface density profiles}
\label{subsec:rot_support}

\begin{figure*}
    \includegraphics[width=\textwidth]{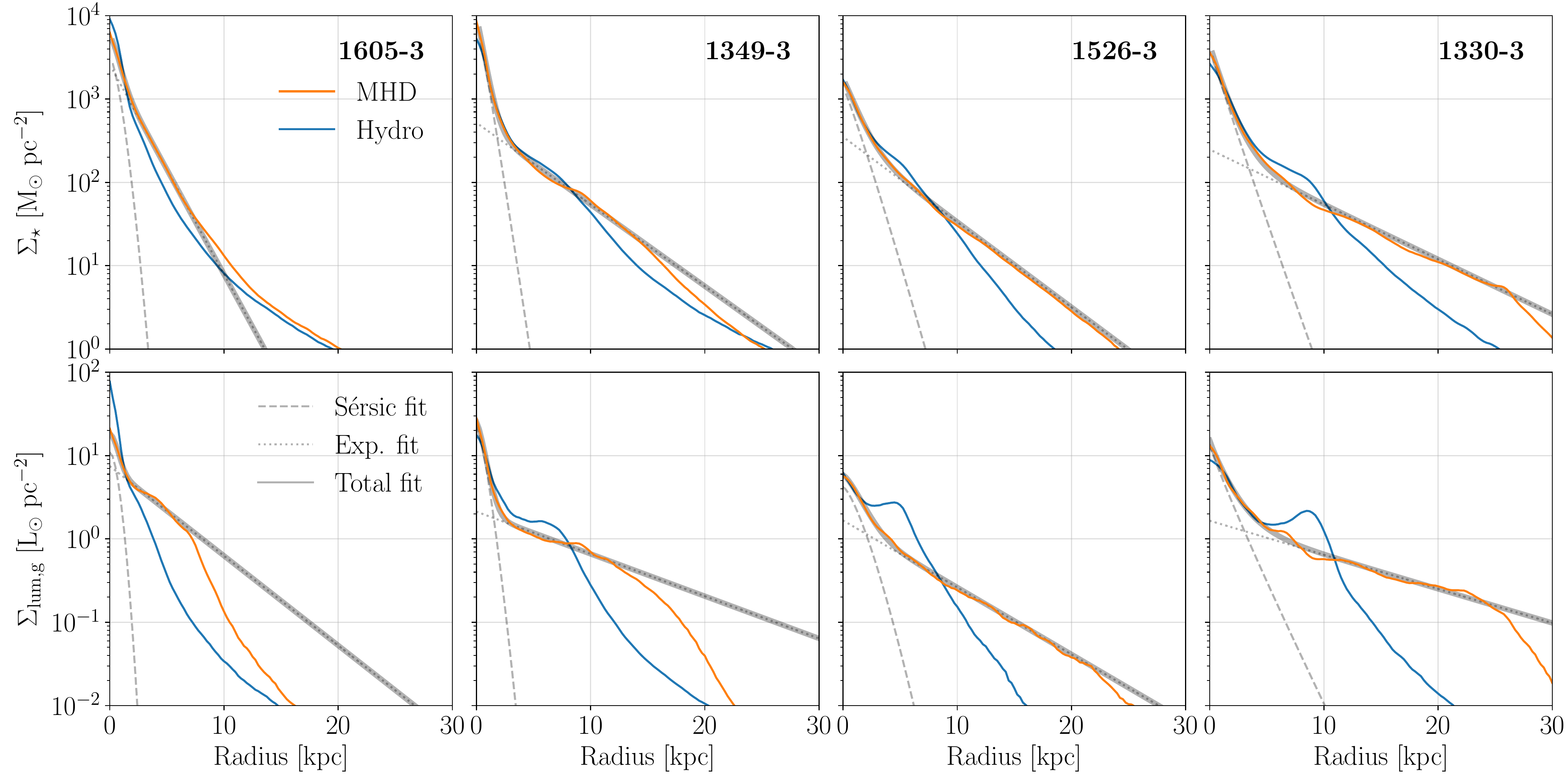}
    \caption{\textit{Top row:} stellar mass surface density profiles for the merger remnant in each high-resolution simulation, as seen at $z=0$ ($z=0.11$ for 1605-3). Profiles are calculated over a height of $\pm5$ kpc from the midplane. \textit{Bottom row:} as above, but showing stellar luminosity surface density profiles for the mock SDSS $g$-band. Data from MHD simulations are fit simultaneously with exponential and S\'{e}rsic profiles using a non-linear least squares method. The functional form of the remnants from hydrodynamic simulations prevents a similar fit, particularly in the case of the luminosity profiles, due to the stellar ring component.}
    \label{fig:profiles}
\end{figure*}

Galaxies are often characterised by their amount of rotational support. We parameterize this here for each remnant using the orbital circularity parameter, as defined in \citet{abadi2003}. This is calculated for each star particle within $R_\text{opt}$ as $\epsilon = j_{z} / j(E)$, where $j_{z}$ is the specific angular momentum of the particle aligned with the $z$-axis, and $j(E)$ is the maximum specific angular momentum possible given the particle's specific binding energy, $E$. As in \citet{grand2017}, we then calculate the fraction of stellar mass that kinematically belongs to the disc through two different methods. For the first method, we assume that the bulge makes up twice the mass of the counter-rotating material, where counter-rotating particles are defined by having $\epsilon < 0$. Subtracting the bulge mass from the remaining stellar mass then provides the disc mass. For the second method, we infer the disc mass by summing the mass of particles with $\epsilon>0.7$.  The disc-to-total stellar mass values are shown for each remnant in Table~\ref{tab:sim_data}, with the first and second methods producing the unbracketed and bracketed numbers, respectively. We also show the inferred stellar disc and bulge masses, as given by the first method, in the same table in columns 7 and 8. For many cases, the disc-to-total mass ratio is not substantially changed between physics models. This is a result of the competing factors that produce this ratio; whilst MHD simulations produce much larger, flatter stellar discs, they also have a propensity to produce remnants with bulges. Furthermore, whilst the remnants in hydrodynamic simulations are smaller, they tend to concentrate stellar mass in a ring, thereby increasing their overall circularity fraction. The picture is further muddied, as larger galaxies are more susceptible to warping effects at the disc edge, brought on by the tidal interactions, which reduce the circularity fraction. Examples of this may be clearly seen in the edge-on mock images for 1526-3M and 1330-3M in Fig.~\ref{fig:mocks}. 

In addition to the disc-to-total values shown in Table~\ref{tab:sim_data}, we also calculate these ratios considering only stars that were born within the last 2 Gyr. By doing so, we isolate the young stellar disc in the remnant and exclude contributions from stars whose orbits were disrupted in the merger. Naturally, the resultant disc-to-total ratios tend to be marginally higher, as some of the more eccentric orbits are removed. However, the overall picture remains qualitatively the same: most disc-to-total values are very similar regardless of physics model used and any changes that do exist are not systematic.

This conclusion is somewhat in contradiction with that arrived at by \cite{vandeVoort2021}, who find that disc-to-total values increase systematically with the inclusion of magnetic fields in the galaxy formation model. To an extent, this incongruity is explained by a difference in definition; instead of orbital circularity, the parameter $\kappa_\text{rot}$ is used \citep[see, e.g.][]{sales2012}, which measures the fraction of kinetic energy in ordered rotation. Calculating this parameter for our own galaxies, we find that two of the four high-resolution simulations (1330 and 1349) show a comparable increase with the addition of MHD physics. The remaining two, however, show very similar values. We note that 1330-3H and 1349-3H have particularly well-pronounced bars and we believe that it is this feature that reduces the $\kappa_\text{rot}$ parameter below that of their MHD analogues. This interpretation is also consistent with the galaxy studied in the main body of \cite{vandeVoort2021}.

From a more observational perspective, the fraction of mass in the bulge and disc components may also be calculated by fitting stellar mass and luminosity surface density profiles with exponential and \citet{sersic1963} profiles. We show these profiles for the remnants from all high-resolution simulations in Fig.~\ref{fig:profiles}, with exponential and S\'ersic profiles overlaid for the MHD simulations. These fits were calculated simultaneously using a non-linear least squares method. Although not shown here, fits were also made for the hydrodynamic stellar mass surface density profiles. The radial scale length, bulge effective radius, and S\'ersic index for the stellar mass surface density profiles for all simulations are given in Table~\ref{tab:sim_data} for comparison. These quantities generally vary little between physics models, and are only systematic in the case of the S\'ersic index, which always has a lower value in the case of the hydrodynamic simulations. This is a result of the more core-like centres for these merger remnants, which most likely results from their stellar bar component.

Data is not provided for the MHD $g$-band luminosity profile fits, but it can be seen by eye that the radial scale length is much longer for the luminosity profile than for the respective mass profile. This is a result of the inside-out growth of the disc, resulting in a younger population on average at the disc edge. For the hydrodynamic simulations, the stellar ring produces an unusual `sombrero' shaped luminosity profile. This shows clearly how star formation has been concentrated in this region. This feature is to be seen to some extent in all of the broad bands available, but is particularly clear in the $g$-band, as presented here in the bottom panel of Fig.~\ref{fig:profiles}. The maxima at the ring component is followed by a dramatic drop in stellar density. This further shows how the expansion of the galaxy has been curtailed. Whilst lenticular galaxies often show some flattening of the luminosity profile at the stellar ring \citep{buta1996}, maxima and sharp drop-offs in density as seen here are highly unusual. The remnants produced in the hydrodynamic simulations are therefore not only poorly described by the standard disc profile fit, but are also atypical of observed galaxies generally.

Whilst better fit than their hydrodynamic counterparts, the luminosity profiles produced in MHD simulations are also not perfectly fit by the standard exponential and S\'ersic profiles. This is because the remnants in our simulations generally consists of a superposition of new and old stellar discs, rather than one unified disc. Furthermore, these discs are both situated on top of a stellar halo, which itself has often been extended and distorted by the merger. The result is that the remnants do not necessarily show the clearly defined up- and downwards-bending breaks needed for an exact fit, and lack a well-defined edge. This means that the integration bounds for calculating disc-to-total ratios using these fits are unclear. Consequently, we refrain from calculating disc-to-total ratios using this method.

In \citet{grand2017}, stellar mass surface density profiles were fit out to $R_\text{opt}$, defined as the radius at which the $B$-band surface brightness drops below $\mu_{B} = 25$ mag arcsec$^{-2}$. Whilst this quantity does not necessarily define the exact edge of the galaxy, as just discussed, it still provides a useful bound on the disc size. In particular, it is able to capture the difference in the disc size produced by each physics model well, and allows for a quantitative comparison between our simulations and the fiducial `Level 4' Auriga simulations. We list $R_\text{opt}$ for all our simulations in the penultimate column of Table~\ref{tab:sim_data}. By comparing the two, it can be seen that the merger remnants from the hydrodynamic simulations are significantly more compact than the fiducial Auriga galaxies. The MHD simulations, on the other hand, produce remnants that are of comparable size to the fiducial galaxies. This is not too surprising as in this case both simulations use the same physics model.

\subsubsection{Impact of MHD for more isolated galaxies}
\label{subsec:isolated_galaxies}

\begin{figure*}
    \includegraphics[width=\textwidth]{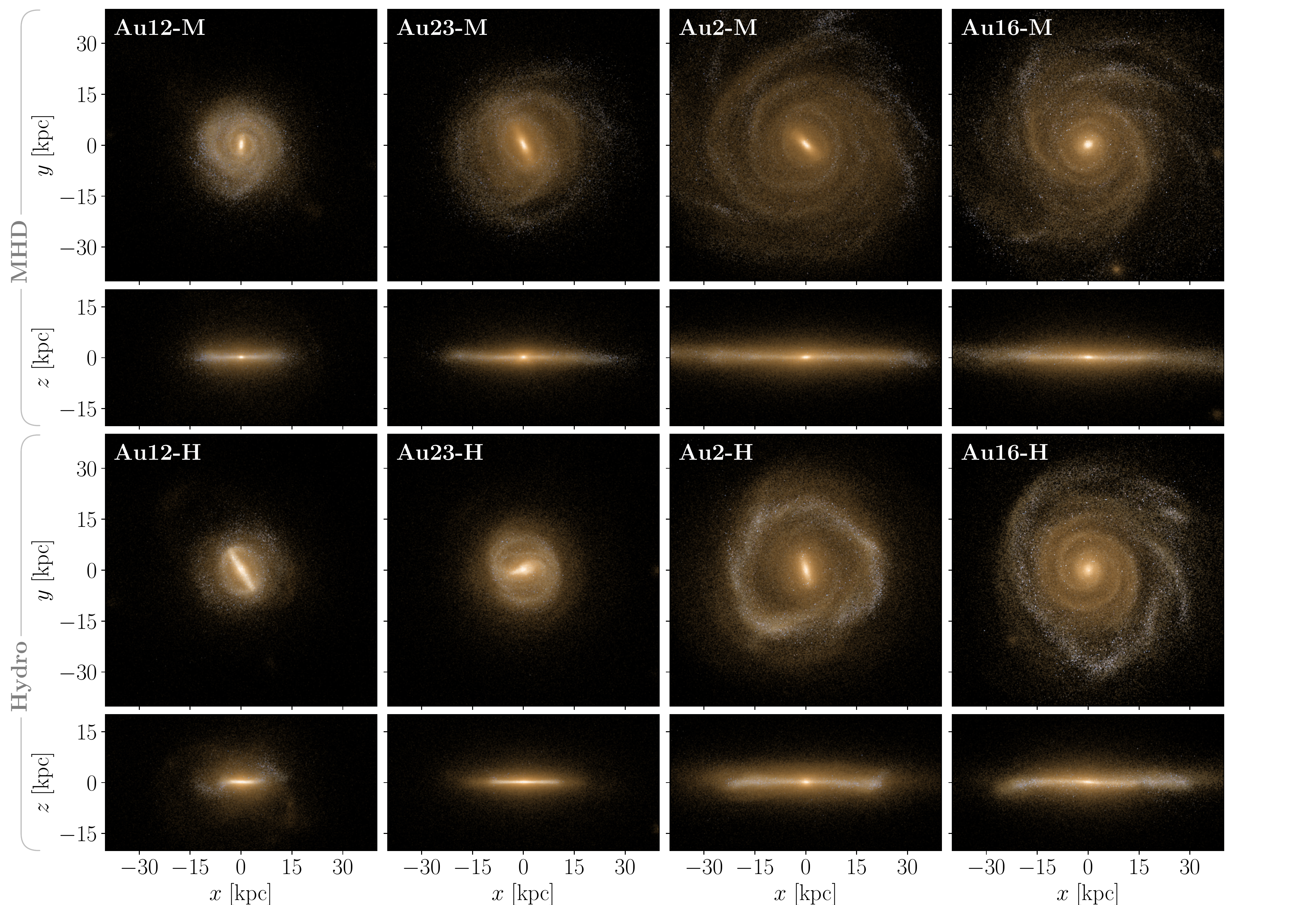}
    \caption{As Fig.~\ref{fig:mocks}, but now showing mock SDSS \textit{gri} composite images for galaxies from the Auriga simulations \citep{grand2017}. The top row continues to show MHD simulations, whilst the bottom row shows their hydrodynamic analogues. Similar morphological features to those seen in Fig.~\ref{fig:mocks} are evident here too, but the differences are less marked. The galaxies presented have all had more quiescent merger histories than those seen in Fig.~\ref{fig:mocks}, but they have not experienced complete isolation. We interpret this as evidence that the features seen are predominantly produced by mergers and that such features are generally stable over time.}
    \label{fig:mocks-auriga}
\end{figure*}

The comparison of our simulations with the fiducial Auriga galaxies also allows us to isolate the role of mergers in producing the observed morphologies. As discussed in Section~\ref{subsec:methods_isolated_galaxies}, four of the Auriga simulations were run using both MHD and hydrodynamic physics models. We present face and edge-on mock SDSS images for each of these in Fig.~\ref{fig:mocks-auriga}, created in the same manner as for Fig.~\ref{fig:mocks}. The simulations retain their name from the fiducial runs, albeit with the addition of an `M' or `H', indicating the inclusion of MHD or hydrodynamic physics, respectively. The simulations are also all `Level 4' in the Aquarius nomenclature \citep{marinacci2014}, with a dark matter mass resolution of $3 \times 10^5 \; \text{M}_\odot$. This mass resolution is almost exactly between that of our highest and intermediate-resolution simulations. Trivially, this lower resolution increases the minimum scale at which physical structure may form. In practice, the difference is not great enough to substantially affect the produced morphology.

At first glance, many of the features evident in our own simulations may also be observed in Fig.~\ref{fig:mocks-auriga}. For example, whilst larger stellar bars are now also seen in the MHD simulations, they are still generally more extended in the hydrodynamic simulations. This is taken to its extreme for Au12-H, where the bar transverses almost the entire length of the disc. On top of this, we continue to see stellar rings in hydrodynamic galaxies, whilst they do not appear in MHD simulations. Even the largest hydrodynamic remnant shows evidence of a stellar ring, despite this ring being fairly distorted. Coincidentally, Au16-H is the only hydrodynamic galaxy that does not display an extended stellar bar. It therefore seems likely that the bar structure is providing the resonant forces that generate and maintain the stellar ring. This follows from theoretical predictions that gas should accumulate at Lindblad resonances, under the continuous action of gravitational torques \citep{buta1996, rautiainen2000}.

\begin{figure*}
    \centering
    \textbf{Resolution study}\par\medskip
    \includegraphics[width=\textwidth]{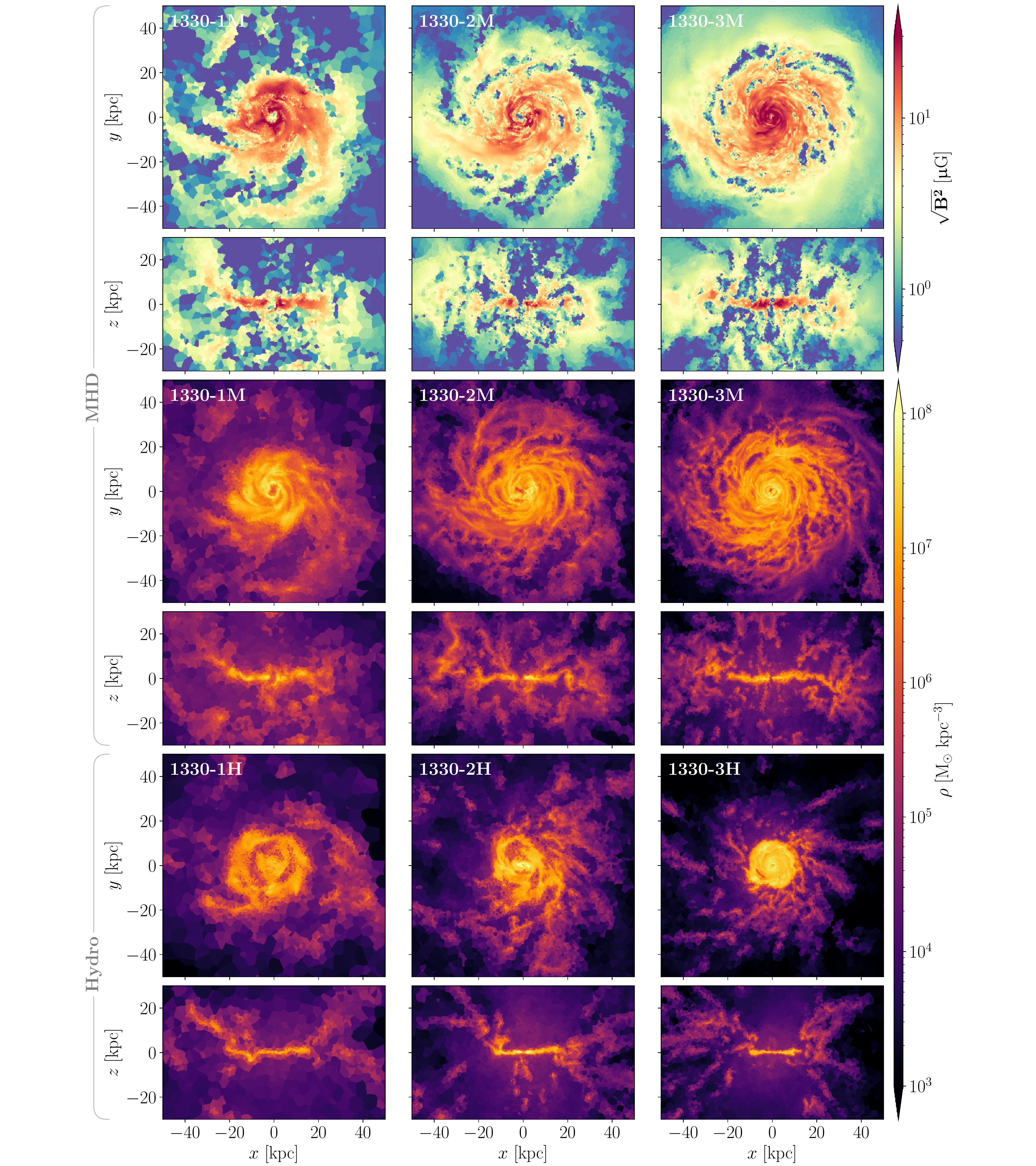}
    \caption{As Fig.~\ref{fig:gas}, but now showing face and edge-on slices through the merger remnant for simulations with increasing resolution (left to right). The magnetic field as a function of radius becomes smoother and more extended with increased resolution but is otherwise very similar. The gas morphologies, on the other hand, show divergent evolution as a function of resolution. In particular, the gas disc in the hydrodynamic simulations becomes thinner and more compact with each increase in resolution, whilst the gas disc in the MHD simulations grows slightly and becomes noticeably more flocculent.}
    \label{fig:gas_zoom}
\end{figure*}

Whilst there are many similarities between the Auriga galaxies and our own simulations, there are also clear differences. For example, whilst galaxies from MHD simulations are still generally larger than those in hydrodynamic simulations, this difference is nowhere near as stark as it was in our own merger simulations. We may quantify this difference by comparing the optical radii, $R_\text{opt}$, of each pair of galaxies. For our merger simulations, the MHD variant is on average 54\% larger than the hydrodynamic analogue. This relative size difference increases as the galaxies become larger. In comparison, for the Auriga galaxies the $R_\text{opt}$ values of the MHD variant are on average only 20\% larger. This difference does not increase with the size of the galaxies. Indeed, both Au2-H and Au16-H have developed quite large discs, relative to those seen in our own hydrodynamic simulations. The differences in interior morphology are also not as clear-cut. For example, Au16-H has also developed spiral arm structure in the central part of the disc, whilst no remnant in our own hydrodynamic simulations was able to form this structure. The gas distribution in a galaxy can, of course, be heavily disrupted by the existence of a stellar ring or bar component, and the weakness of these features in Au16-H has likely allowed the spiral structure to form here.

In comparing the more isolated galaxies with our own merger remnants, we conclude that the morphological differences are greatest when the merger history is most active. However, we must also bear in mind that the Auriga galaxies are not perfectly isolated. These simulations, too, are cosmological, and the galaxies are only selected to be isolated from significant tidal interactions at late times. Due to the hierarchical growth of structure in $\Lambda$CDM, these galaxies have naturally undergone mergers at earlier times in their history \citep{bustamante2018, monachesi2019}. We therefore propose that the morphological differences seen in Fig.~\ref{fig:mocks} and Fig.~\ref{fig:mocks-auriga} are primarily a result of MHD effects excited by mergers. The full mechanism for this will be explored in greater detail in an upcoming paper.

\subsection{Resolution study}
\label{subsec:resolution-study}

\subsubsection{Divergent gas morphology}

\begin{figure}
    \centering
    \includegraphics[width=\columnwidth]{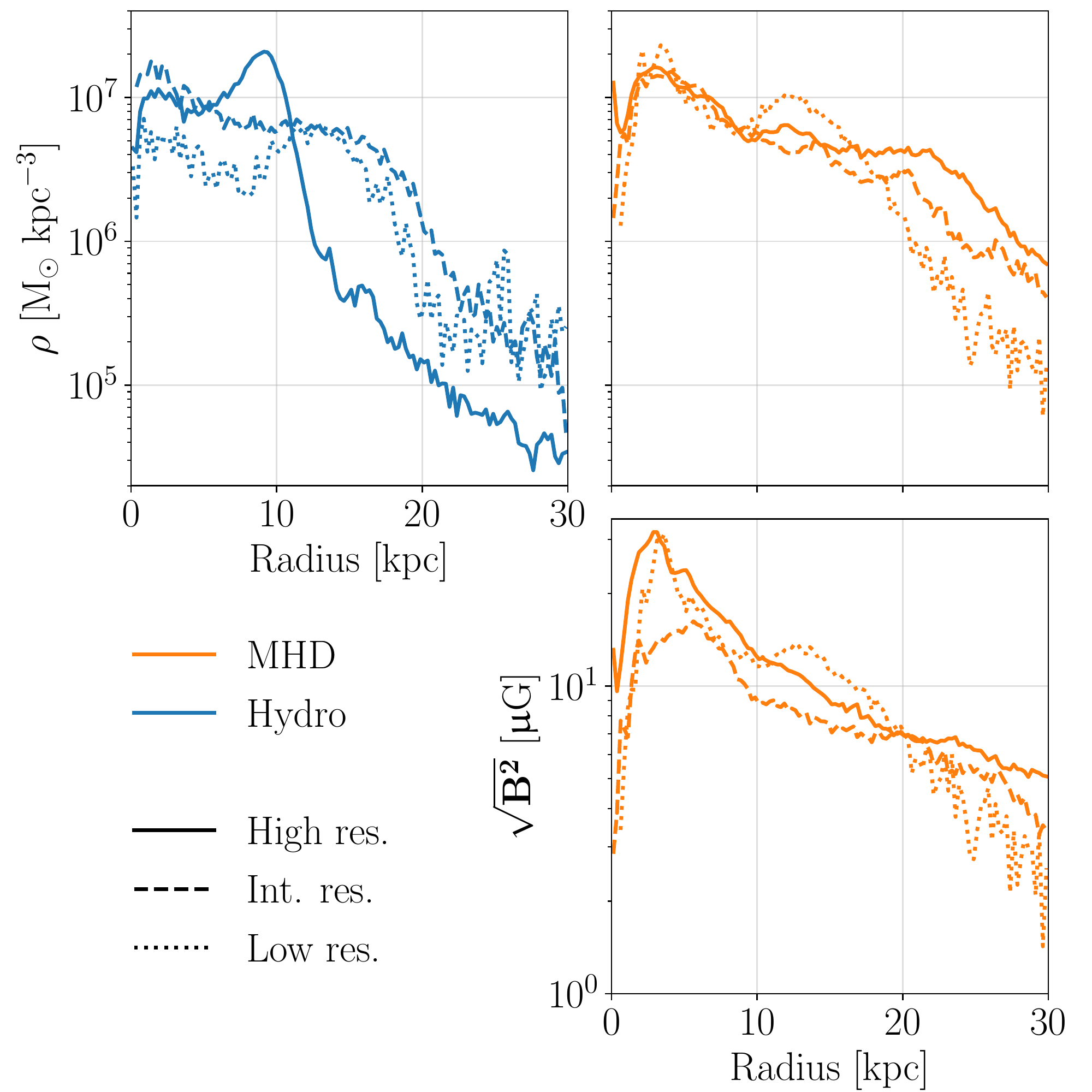}
    \caption{\textit{Top row:} mean gas density as a function of radius for the merger remnants seen in Fig~\ref{fig:gas_zoom}. \textit{Bottom row:} mean magnetic field strength for the MHD simulations of the same figure. In both cases, the mean is calculated over a height of $\pm$5 kpc from the midplane. The gas disc grows in the MHD simulations whilst shrinking in the hydrodynamic simulations.}
    \label{fig:profiles-zoom-gas}
\end{figure}

\begin{figure*}
    \centering
    \textbf{Resolution study}\par\medskip
    \includegraphics[width=\textwidth]{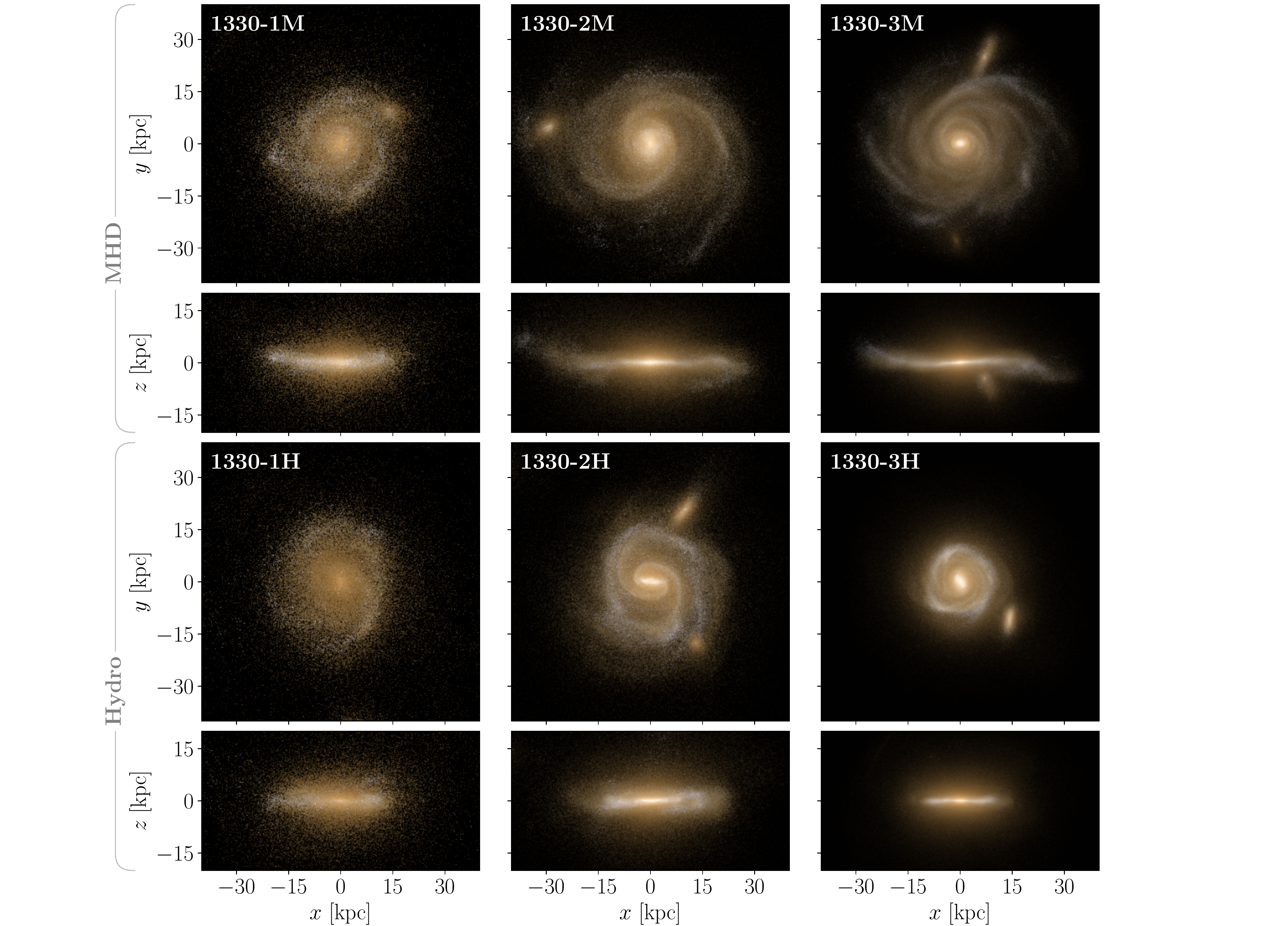}
    \caption{As Fig.~\ref{fig:mocks}, but now showing SDSS \textit{gri} composite mock images for simulations with increasing resolution (left to right). In order to conserve the average luminosity per bin, the images of the lower resolution simulations are progressively coarsened. The lowest-resolution simulations produce very similar merger remnants for both MHD and hydrodynamic models. However, by the intermediate-resolution, strong morphological differences are already apparent. This divergence continues with increasing resolution.}
    \label{fig:mocks-zoom}
\end{figure*}

In addition to our eight high-resolution simulations, we have also run two intermediate and two lower-resolution simulations. In Section~\ref{subsec:MHD-global-properties}, we showed that the global properties of the galaxies as a function of resolution were well-converged. However, as for the high-resolution simulations, this does not quite tell the whole story. Some indication of this may already be observed in Table~\ref{tab:sim_data}, where it can be seen that the optical radius of the remnants is remarkably similar for the lower resolution runs for both MHD and hydrodynamic models. This implies that the structure of the remnants may not be converged with resolution. To investigate this effect, we repeat the analysis performed in Fig.~\ref{fig:gas} for our lower resolution simulations. In Fig.~\ref{fig:gas_zoom}, we show slices taken face and edge-on through the main galaxy in each of the 1330 simulations. Each remnant is seen at $z=0$. Once again, in the top two rows we show the magnetic field strength in each gas cell, whilst in the bottom four rows we show the gas density for MHD and hydrodynamic simulations, respectively. We quantify the results at this time by showing radial profiles of these properties in Fig.~\ref{fig:profiles-zoom-gas}.

The magnetic field profiles continue to be generally axisymmetric at lower resolution, but the radial gradient of the field strength is not as smooth. There are also some clear differences in the structure of the magnetic field outside the central 10 kpc. For example, the lowest-resolution simulation shows a rather sharp  decline in its radial profile beginning at approximately 12 kpc, reflecting the decline in gas density starting at the same radius. This effect is somewhat exaggerated by the slight rise in both the mean gas density and mean magnetic field strength shortly before this point, as seen in the right-hand panels of Fig.~\ref{fig:profiles-zoom-gas}. The strength of the magnetic field beyond 20 kpc is also lower and more erratic for the lower resolution simulations; whilst 1330-3M shows a smooth field beyond this point with an average strength of a few $\upmu$G, this is not seen in 1330-1M. The development of a smooth magnetic field at the outer edges of the disc in our highest resolution simulation likely requires the filamentary gas structure seen for this simulation in Fig.~\ref{fig:gas_zoom}. Such structure provides a sufficient average gas density to maintain the field strength, but also provides enough small-scale structure for the gas to be gravitationally bound both to itself and the disc, preventing the density cut-off seen in the lowest-resolution simulation. The development of such small-scale structure, in turn, clearly requires sufficiently high-resolution. As well as affecting structure development, the increased resolution is also likely to be the cause of the smoother, more axisymmetric field profile in the disc in 1330-3M. For this simulation, the higher resolution allows for the gradient in strength between cells to be better resolved, smoothing the resultant magnetic topology. 

Whilst an increase in resolution allows for a smoother gas density and magnetic field strength profile at the disc edge, it also helps a more heterogeneous density distribution to develop above and below the disc. This property is vital for allowing smaller cloudlets to survive the stellar wind, thereby promoting the action of a small-scale fountain flow. As noted earlier, fountain flows of this kind have been found to be crucial for the radial growth of discs in the Auriga model. The supply of gas to the outer edges in this manner may well be supporting the maintenance and growth of the filamentary gas structures at this radius as well.

Considering the disc itself, the radial density profile of the inner 10 kpc is fairly well converged for all MHD simulations as a function of resolution, as can be seen in Fig.~\ref{fig:profiles-zoom-gas}. This is not the case for the hydrodynamic simulations, which show significant variation in their profiles. With this said, there are still clear morphological similarities to be seen for all the hydrodynamic simulations in Fig.~\ref{fig:gas_zoom}. In each case, gas is seen to accumulate predominately at the centre and at the disc edge, with a sharp cut-off in density thereafter. The region above and below the gas disc is also notably lower in density for all resolutions compared to the MHD simulations. The clearance of gas from this region becomes more effective with increased resolution as the disc becomes smaller and the stellar wind becomes stronger. This also reduces the thickness of the disc.

\subsubsection{Divergent stellar morphology}

The differences between the remnants as a function of resolution become even more clear when we inspect the stellar morphology. In Fig.~\ref{fig:mocks-zoom}, we show SDSS \textit{gri} composite mock images for the 1330 simulations, created in the same manner as for Fig.~\ref{fig:mocks}. The top two rows show simulations that include MHD physics, whilst the bottom two rows show simulations that include only hydrodynamic physics. In order to roughly conserve the average luminosity per bin, we have adjusted the bin size relative to the spatial resolution.

Whilst some differences could be seen in the gas structure at the lowest-resolution, the stellar morphologies at the lowest-resolution for both MHD and hydrodynamic models look very similar. Indeed, most morphological differences here can be mostly explained by the slight variations in their respective star formation histories. The radius and scale height of both lowest-resolution remnants are practically identical. The galaxies are notably more puffed up than for the higher resolution runs, but this is predominantly a result of the higher softening length used in our lower resolution simulations (see Table~\ref{tab:sim_setup}), combined with the dynamic nature of the systems. This is supported by the observation that the scale height decreases for both sets of simulations with each increase in resolution.

Whilst the lowest-resolution remnants appear very similar to one another, significant differences are already apparent at the next intermediate-resolution level. In this case, the mass resolution is eight times higher and the morphologies have diverged substantially from one another. In fact, many of the morphological differences observed in Fig.~\ref{fig:mocks} may be seen here too, but on a larger spatial scale. For example, the remnant in the MHD simulation has formed a bulge-like centre with two well-defined spiral arms connecting  to it. These fade with distance from the centre, with the galaxy showing an overall shallow radial luminosity gradient. In contrast, 1330-2H shows a bright stellar ring, with two lanes of stars connecting to a large central bar. When viewed together with Fig.~\ref{fig:gas_zoom}, it is clear that these features are strongly distorting the gas morphology. In particular, the bar coincides with the peak in the gas density, showing how it is drawing gas in from elsewhere in the disc. The accumulation of gas in this manner implies that the gas dynamics here are relatively calm. This is in contrast with the signatures of AGN outbursts observed for the remnants from the MHD simulations, as well as the particularly chaotic gas dynamics seen for 1605-3M and 1526-3H in Fig.~\ref{fig:gas}. The morphology seen for 1330-2H in Fig.~\ref{fig:mocks-zoom} is perhaps the most clear evidence seen yet that the ring morphology produced in the hydrodynamic simulations is generated from bar-driven orbital resonances. The full confirmation of this effect is, however, left to an upcoming paper. 

It may also be observed that the ring morphology is fairly robust; in the face-on panel for 1330-2H, it can be seen that the remnant is being harassed by two small satellite galaxies. Despite this, the ring morphology is still very much intact. The reinforcement of the morphology through the accumulation of gas under gravitational torques, as well as the robustness of the morphology to small gravitational perturbations, provides a further indication that the features noted are durable. This helps to explain their appearance in the more isolated galaxies analysed in Fig.~\ref{fig:mocks-auriga}. 

\begin{figure}
    \centering
    \includegraphics[width=\columnwidth]{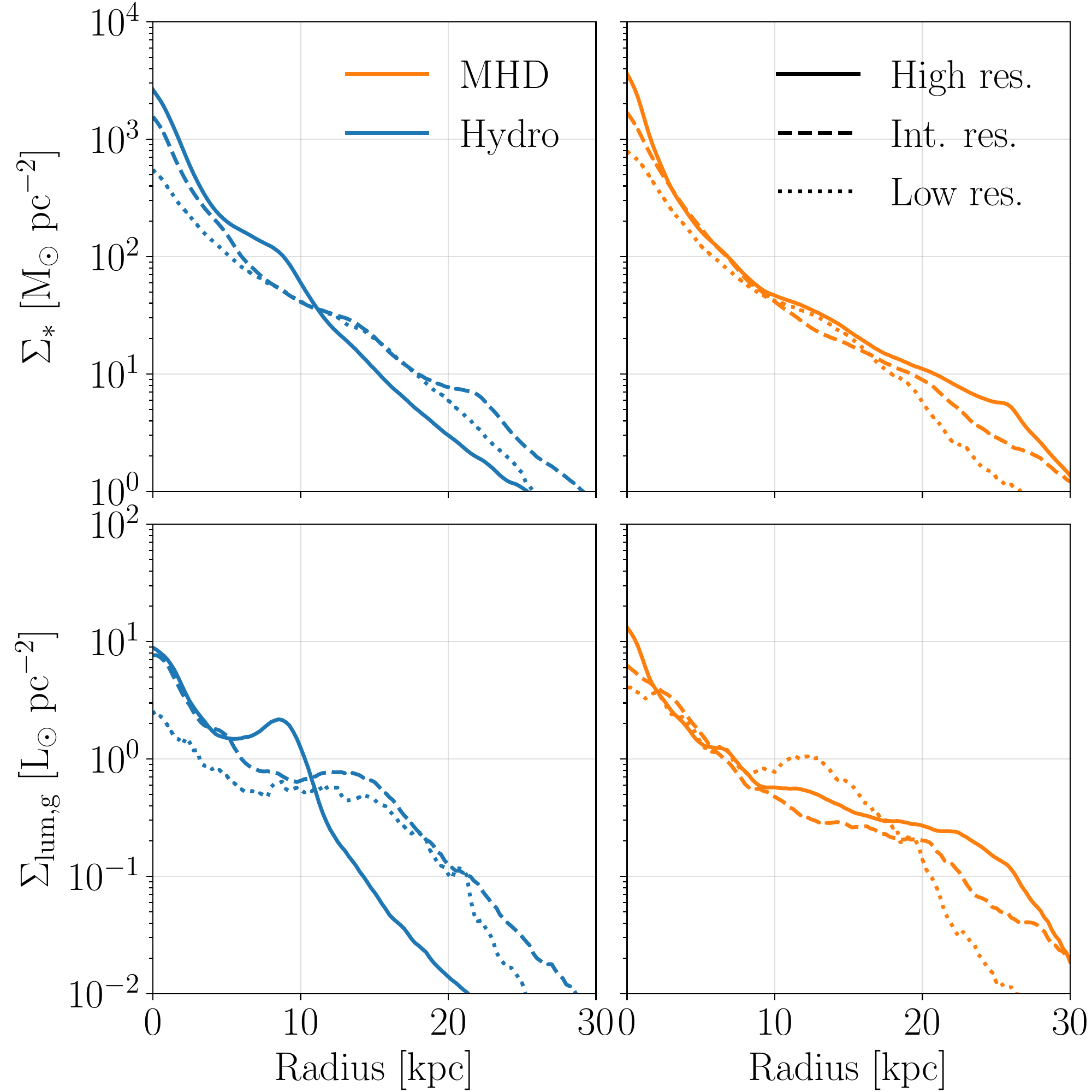}
    \caption{As Fig.~\ref{fig:profiles}, but now showing stellar mass and luminosity surface density profiles for the 1330 simulations. Once again, the profiles produced by each physics model are very similar for the lowest-resolution, but diverge with increasing resolution.}
    \label{fig:profiles-zoom}
\end{figure}

In Fig.~\ref{fig:profiles-zoom}, we examine the differences in the structure more quantitatively. As in Fig.~\ref{fig:profiles}, we show the stellar mass surface density in the top row and stellar luminosity surface density in the mock SDSS $g$-band in the bottom row. The hydrodynamic 1330 simulations are shown on the left, whilst the MHD 1330 simulations are shown on the right. As expected, the lowest-resolution profiles are very similar. Indeed, the MHD simulation even shows evidence of a stellar ring in the luminosity profile, which was not wholly clear in the mock images. As the resolution increases, this disappears, and the remnants in the MHD simulations take on a classic disc galaxy profile. This becomes more extended with resolution, but does not change substantially in form. A slight increase in the stellar mass in the inner regions is seen with increased resolution, as was previously discussed in Section~\ref{subsec:MHD-global-properties}. This increase is also seen for the hydrodynamic simulations. Despite the significant development in the stellar morphology seen between the low and intermediate resolution hydrodynamic simulations in Fig.~\ref{fig:mocks-zoom}, the overall radial stellar surface density profiles remain broadly similar for both physics models for the inner $\lesssim$20 kpc. The luminosity profiles, too, are not drastically different. Indeed, the full `sombrero'-style profile only develops at the highest-resolution hydrodynamical simulation. Whilst this results from a range of factors, the two key factors are likely to be the smaller softening length and slightly higher star formation rate in the highest resolution simulations. Together these factors allow for a high star-formation density, which is able to launch a strong stellar wind.

\subsubsection{Resolution study of the magnetic dynamo}
\label{subsec:amplification_res}

\begin{figure*}
    \includegraphics[width=\textwidth]{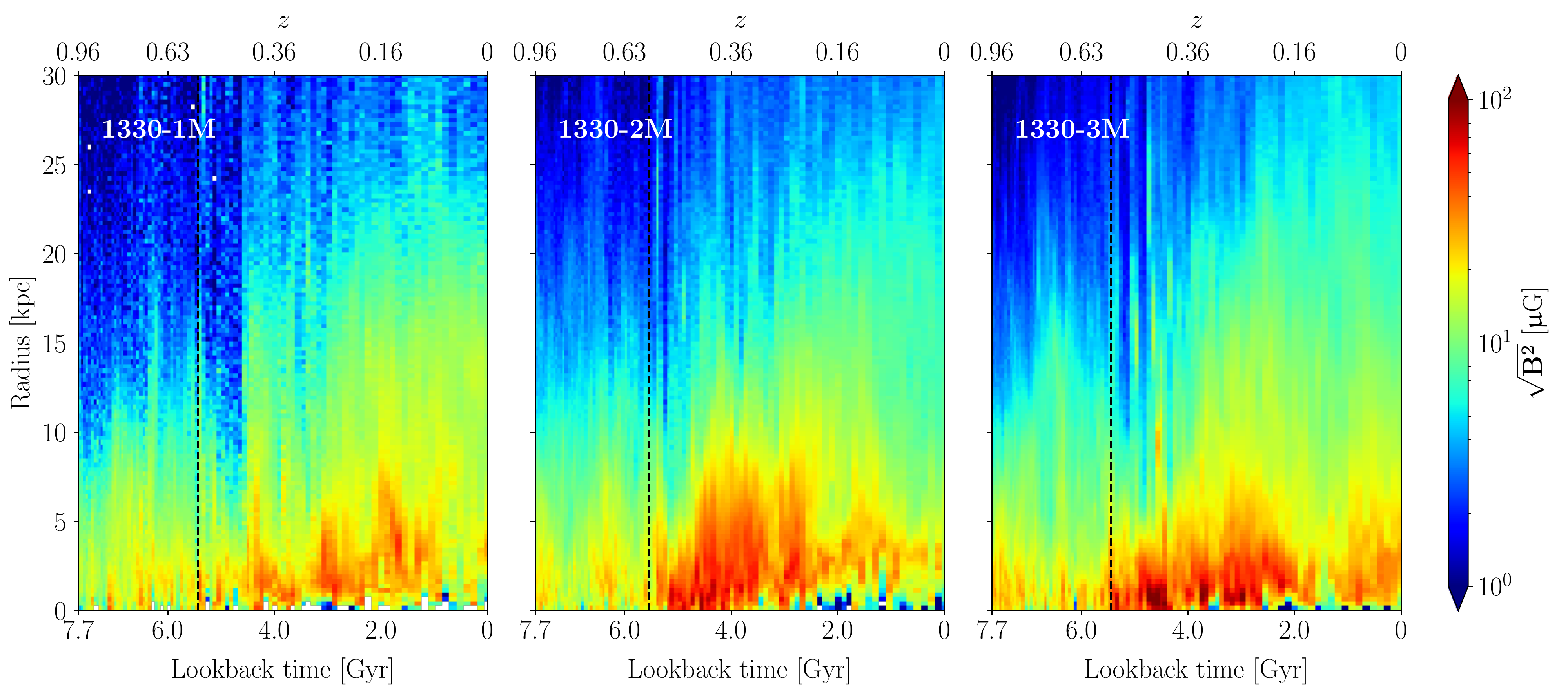}
    \caption{The radially-binned mean magnetic field strength, as in Fig.~\ref{fig:BH_mag}, but for simulations with increasing resolution (left to right). The dashed vertical line marks the time of first periapsis. Whilst the outer reaches of the galactic disc show relatively well-converged evolution, the amplification in the inner regions is substantially stronger for higher resolution simulations. The resulting magnetic field strength is better able to affect the gas flow, ultimately producing a different gas and stellar morphology.}
    \label{fig:mag_zoom}
\end{figure*}

\begin{figure*}
    \includegraphics[width=\textwidth]{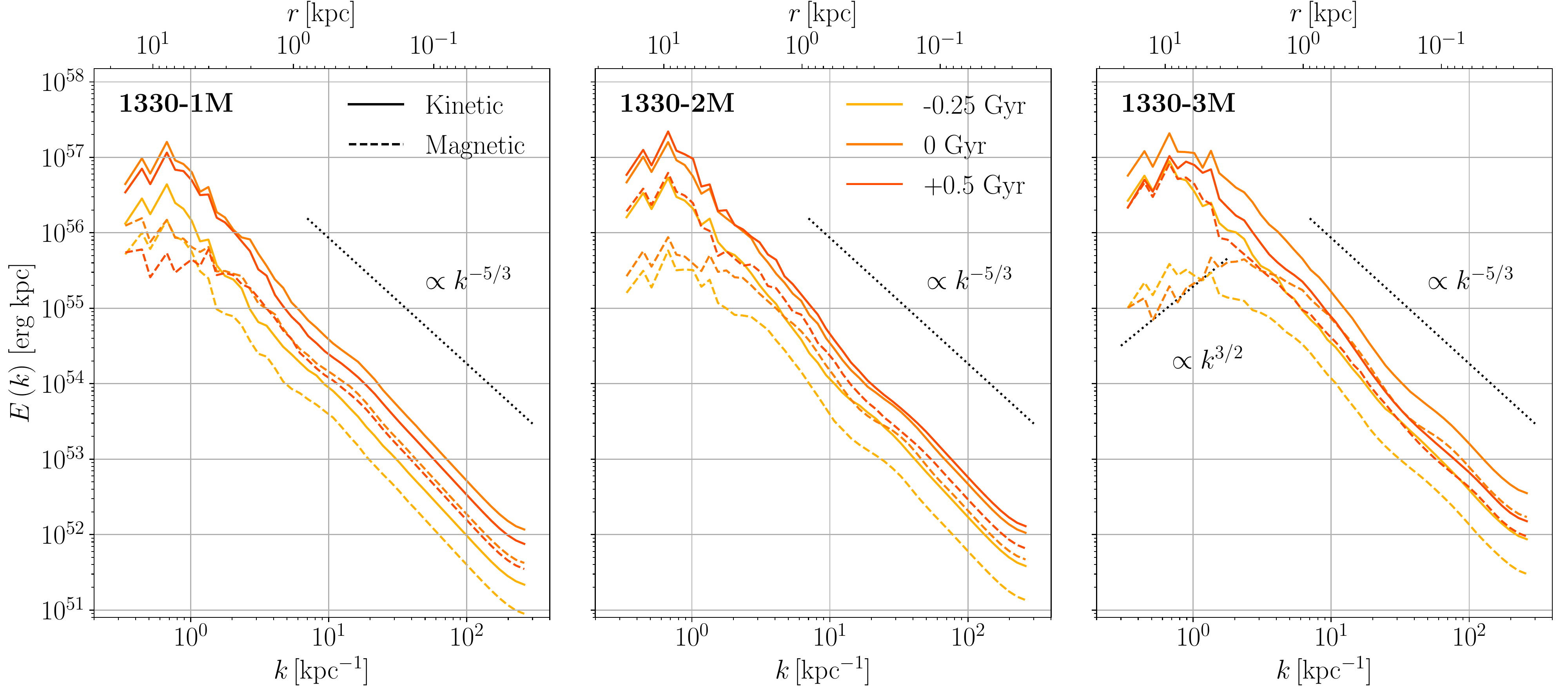}
    \caption{Kinetic and magnetic energy power spectra for the 1330-M simulations, calculated using all gas cells within 5 kpc of the galactic centre. Times are shown from first periapsis ($t=0$~Gyr). The black dotted lines show the slopes of a \citet{kolmogorov1941} spectrum ($\propto k^{-5/3}$)  and a \citet{Kazantsev1968} spectrum ($\propto k^{3/2}$), which are theoretically expected for a small-scale dynamo resulting from incompressible turbulence. Whilst the kinetic energy power spectrum initially evolves in a similar fashion for all simulations, the amplification of the magnetic field is more efficient with increased resolution, as seen by the increase in magnetic energy at larger scales over time.}
    \label{fig:power_spec}
\end{figure*}

As resolution increases, the galaxies transition towards a particular morphology. This transition is consistent, as seen by the regularity with which the characteristic morphological features form in the highest-resolution simulations. We therefore argue that the divergence of morphology with resolution points to the importance of including small-scale physics in the simulations, rather than towards general numerical divergence. This claim is bolstered by the global properties of the remnants, which, as previously shown, are broadly converged across all resolutions and physics models. Whilst the ability to form increased amounts of small-scale structure with higher resolution has a non-trivial impact on the remnant development, the impact of the magnetic fields themselves becomes stronger with increased resolution too. A study of this will help us understand how the morphology forms and whether we should expect even further divergence with still higher resolution.

In Fig.~\ref{fig:mag_zoom}, we show the radially-binned mean magnetic field strength as a function of time for simulations of different resolution. This was calculated in the same manner as discussed for Fig.~\ref{fig:BH_mag}. The general radial evolution of the field in the outer reaches is well-converged, showing a similar strength for all times. The inner regions, however, are clearly more strongly amplified for the higher resolution simulations; whilst the lowest-resolution simulation rarely shows mean field strengths higher than 30 $\upmu$G, strengths regularly reach between $40-50$ $\upmu$G in the intermediate-resolution simulation, and upwards of 70 $\upmu$G in the highest-resolution simulation. Such strengths massively increase the ability of the magnetic field to affect the local gas dynamics. This is particularly so shortly after periapsis, when the field strengths are highest, and the rebuilding of the stellar disc has already begun. The evolution of the merger remnant during this time is crucial to its further development, and so the impact of the increased amplification here is itself magnified.

The origin of this increased amplification lies, almost certainly, in the more efficient excitation of the small-scale dynamo. This can be seen by inspecting the evolution of the magnetic and kinetic power spectra around the time of the initial injection of turbulence. We show this for the three different resolution simulations of the 1330 galaxy model in Fig.~\ref{fig:power_spec}. Following \citet{pakmor2011, pakmor2017}, we compute these power spectra by taking the absolute square of the Fourier transforms of the components of $\sqrt{\rho}\,\bs{\varv}$ and $\bs{B}/\sqrt{8\pi}$, respectively \citep[cf.][]{bauer2012}, for gas within a sphere of radius 5 kpc. This is done within a zero-padded box of size $\pm$10 kpc across, and therefore the fundamental mode has a wavelength of 20 kpc. Drops in the power spectra on scales greater than 10 kpc are an artefact of this zero-padding. By considering only gas that lies within 5 kpc of the galactic centre, we isolate the region in which the greatest amplification takes place. We show a time progression from just before the first periapsis, when most of the turbulence is injected, until shortly afterwards. These results do, however, hold for a range of times and radial cuts. In addition, we have also looked at the power spectra of \textit{specific} energies -- both kinetic and magnetic. By examining these, we confirm that only a small part of the evolution can be explained by adiabatic compression.

In each case, the approach of the merging galaxy increases the total kinetic energy in the volume by a factor of roughly 3.5; far above the usual fluctuations. This results in a shift of the kinetic power spectra upwards. The rate of injection of turbulence is sufficiently large such that for a short time afterwards the kinetic energy dominates over the magnetic energy (as was seen previously in Fig.~\ref{fig:BH_mag}). In this regime, the magnetic field is expected to grow exponentially on the corresponding eddy turnover scale, with a growth rate of $\Gamma_l \sim \varv_l / l$, where $\varv_l$ is the eddy velocity at scale $l$ \citep{subramanian1998}. Subsonic, incompressible turbulence, as expected in our mergers, will produce a Kolmogorov-like spectrum of $E(k) \propto k^{-5/3}$ in the inertial range, with $\varv_l \propto l^{1/3}$ \citep{kolmogorov1941}. Together, this results in a growth rate of $\Gamma_l \propto l^{-2/3}$, which means that the magnetic field grows fastest on the smallest scales.

Once the magnetic field becomes strong enough to have a significant dynamical back-reaction at this scale, the exponential growth phase ends and the non-linear growth phase begins \citep{schleicher2013, schober2013}. Under the strong turbulence limit, the magnetic energy in the inertial range then follows the relations: $E(k) \propto k_\perp^{-5/3}$ and $k_\parallel \propto k_\perp^{2/3}$, where $k_\parallel$ and $k_\perp$ are the components of the wave number parallel and perpendicular to the mean magnetic field \citep{goldreich1995}. At smaller scales, such anisotropies are averaged over in our power spectra, producing both kinetic and magnetic energy spectra that follow the perpendicular scaling of $k^{-5/3}$ \citep[see, e.g.][]{beresnyak2019}. For this scaling, the total magnetic energy grows linearly in time \citep{schober2013}. As it saturates at smaller scales, the peak of the magnetic energy spectrum shifts to ever larger scales  (i.e. smaller $k$). At scales larger than this peak, the kinetic energy still dominates, resulting in the familiar Kazantsev slope of $k^{3/2}$ from a kinematic dynamo \citep{Kazantsev1968}. The most clear example of this slope in Fig.~\ref{fig:power_spec} is for 1330-3M, where the difference between the peak scales of the magnetic and the kinetic energy power spectra is greatest.

The kinetic energy itself peaks at the driving scale (or energy injection scale) of the turbulence \citep{cho2009}. This is a factor of a few larger than the greatest scales shown in Fig.~\ref{fig:power_spec}. The amplitude of the kinetic power spectra will decay with time after periapsis, and the spectra as a whole can be affected by sufficiently strong magnetic tension. This is seen once again in 1330-3M, where a strongly saturated magnetic field has shifted the kinetic energy spectrum downwards in the final time step. Indeed, it is only at this highest resolution that the magnetic field is able to saturate to this extent. Despite the initially similar evolution of the kinetic energy budget, it is clear that the magnetic energy evolution at larger scales varies strongly with resolution level. We believe this variation is a direct result of the different growth rates in each simulation; the higher resolution simulations have a lower average cell size, allowing us to resolve smaller eddies. As these eddies have a faster turnover time, the magnetic field saturates sooner at the smallest scale, allowing for the earlier onset of the non-linear growth phase. 

Such differences in growth rates did not affect the fiducial Auriga galaxies, as here the turbulent driving took place over the duration of the galaxy’s initial assembly, being likely a result of cosmic filamentary accretion and stellar feedback \citep{pakmor2017}. In this scenario, the magnetic field could saturate even in lower resolution simulations, as it was given sufficient time in which to do so. In contrast, in our simulations the turbulence is driven by the merger, and the driving time of the turbulence is therefore short compared to the growth and saturation time-scale of the magnetic field. The upshot of this is that in our lower resolution simulations, the field grows too slowly to be able to saturate in the given time frame. The consequence of this can be seen in the time progression for 1330-1M in Fig.~\ref{fig:power_spec}. Here, an increase in the kinetic energy available post-interaction leads to a decrease in the magnetic energy at $k \lesssim 2\,\rmn{kpc}^{-1}$. This happens as the dynamo is unable to saturate quickly enough at higher $k$ values given the new kinetic energy available. In contrast, 1330-2M is able to saturate at the smallest resolved scale sooner, allowing for magnetic energy to cascade to larger scales in time. The result is that the magnetic energy at $k \lesssim 2\,\rmn{kpc}^{-1}$ grows significantly. This process proceeds even more quickly for 1330-3M. In Fig.~\ref{fig:power_spec-high-res}, we show that this behaviour is true of all our high-resolution simulations.

The behaviour of the magnetic field on longer time-scales is more non-linear, as the amplified magnetic fields become better able to impact the gas dynamics and the resultant kinetic energy power spectrum. This leads to fluctuations in the strength of the magnetic field, as can be seen in Figs.~\ref{fig:BH_mag} and \ref{fig:mag_zoom}. The increased speed with which the small-scale dynamo acts in higher resolution simulations, however, helps it to respond to these fluctuations, allowing it to maintain higher magnetic field values over time. Such values are maintained for as long as the kinetic energy is available.

Due to the already high level of saturation reached in the highest-resolution simulations, we do not expect that a further increase in resolution would lead to another significant increase in the average field strength. Higher resolution would, however, result in the yet quicker completion of the exponential growth phase, which would allow the magnetic fields to start affecting star formation earlier. This will continue to be the case until the dissipation scale is resolved. It is unclear to what extent this additional influence would further alter the morphology of the remnants. We expect, though, that other resolution-dependent effects would soon become as, if not more, important, some of which we discuss in the following section.

\section{Discussion}
\label{sec:discussion}

\subsection{Why magnetic fields have been ineffectual in previous simulations}

It is clear from the previous subsection that the influence of magnetic fields in the simulations is strongly dependent on resolution. It is also clear that the morphological differences produced by the two physics models are most distinct after a major merger. These points alone explain why previous simulations run with lower resolution, as well as simulations of isolated galaxies, have not observed a similar impact from the inclusion of MHD physics. There are, however, also other factors at play. For example, we note that our simulations included a comprehensive feedback model, including explicit AGN and stellar wind subgrid models, which may have provided a supplementary role in generating turbulence and certainly affected the accretion of gas. In contrast, the idealised MHD merger simulations that have gone before us only included implicit stellar feedback or included no feedback models at all. This is problematic, both in terms of correctly amplifying the magnetic field \citep{martin-alvarez2018, su2018} and on a more general level, as explicit inclusion of feedback has been shown over the last few years to be a crucial step to generating realistic galaxies \citep{hopkins2014, hopkins2018, marinacci2019}. 

Furthermore, we note the importance of reproducing the correct magnetic field strength as a function of radius, especially in the progenitors. As the magnetic energy density increases with $|\bs{B}|^2$, the field strength must only be lowered by a factor of a few before it becomes subdominant again, as was seen in \citet{hopkins2020}. Observations by the next generation of radio telescopes -- e.g. MeerKAT, SKA, LOFAR \citep{haverkorn2019} -- will hopefully be able to place more precise bounds on such radial profiles. A key test for simulations will be to match the Faraday rotation data from observed galaxies. Such a comparison was shown explicitly for our own MHD implementation in \citet{pakmor2018} for the Auriga galaxies.

\subsection{Relevance of this work to general galaxy evolution}

As stated in Section \ref{sec:methodology}, our merger scenarios were specifically selected in order to produce interactions where the magnetic fields could have their greatest influence. Therefore, whilst magnetic fields have had a significant impact in these simulations, it is not expected that they should have a significant impact in \textit{every} merger scenario. For example, smaller, more gas-poor progenitors would have weaker initial field strengths, and would be less likely to be able to generate the turbulence necessary to sufficiently amplify the galactic magnetic field. This logic also applies to minor mergers, which would have a less disruptive effect on the main galaxy generally. Furthermore, our results do not apply to the `traditional' merger scenario, where gas is expelled from the galaxy post-merger. Here, there is unlikely to be sufficient time for the magnetic fields to influence the development of the merger remnant before star formation is quenched. The role of magnetic fields in these type of mergers is yet to be determined, but it likely plays a weaker part. With this said, we note that the fraction of mergers that are both major and gas-rich only increases with increasing redshift \citep{hopkins2010, man2016}. It is therefore possible that magnetic fields have a more general impact on galaxy evolution, even if they do not play a strong role in particular types of merger. Indeed, as shown in Section~\ref{subsec:isolated_galaxies}, the impact that magnetic fields had in early mergers can be felt several Gyr later even in galaxies that have evolved relatively secularly since.

\subsection{Caveats of the work}
Whilst we believe that our results are numerically robust, there are nevertheless caveats to this work regarding physical fidelity. Firstly, we justify the use of the ideal MHD approximation in our simulations as the magnetic Reynolds number for galaxies, which characterises the relative importance of induction to diffusivity, is expected to be on the order of $\sim10^{18}$ or higher \citep{brandenburg2005}. Without magnetic diffusivity, however, the field topology would be invariant. This would be particularly problematic for simulating a mean-field dynamo, which is believed to reorder the large-scale field to become azimuthally-dominant in the disc on a time-scale of $10^8 - 10^9$ years \citep{shukurov2006}. Non-ideal MHD effects such as reconnection, which acts as a source of magnetic energy loss and potentially also as a source of heating in the galactic halo \citep{raymond1992}, would also be neglected. Whilst we do not explicitly account for magnetic diffusivity in our MHD implementation, some effects will nevertheless be modelled incidentally as a result of the inherent \textit{numerical} diffusivity in our simulations. Indeed, due to our limited resolution, the numerical diffusion in our simulations will be stronger than the physical diffusion. Getting below this scale is still well out of reach of galaxy simulations, and so the use of resistive MHD codes \citep[e.g.][]{marinacci2018} in such simulations in the near future will not be possible. Whilst the numerical diffusivity remains stronger than the physical diffusivity, non-ideal MHD effects could, however, be implemented in future work by using subgrid models, such as those employed by \citet{hanasz2009} in their simulations of isolated disc galaxies. This would increase the physical fidelity of the magnetic diffusion process, although the end result is unlikely to significantly affect the outcome of the simulated dynamos.

Some resolution-dependent MHD effects will also have been neglected in our simulations. For example, idealised MHD simulations have found that magnetic fields are able to stabilise cold streams as they pass through the circumgalactic medium \citep{berlok2019}. Such cold streams would alter the accretion history of the galaxy, and could be particularly important for star formation at high-redshift \citep{keres2005}. Magnetic fields have also been found to be able to support the growth of gas clouds above a critical size in hot winds \citep{sparre2020}. This has important ramifications for the multiphase nature of the circumgalactic medium, which further affects galaxy evolution. Both of these effects, however, once again either require either a substantial increase in resolution or the introduction of new subgrid models.

Without increasing resolution, a substantial step forward in physical fidelity could be taken by implementing cosmic ray physics in our simulation \citep{Pfrommer2017}. Cosmic rays are expected to have a comparable energy density to magnetic fields in the ISM \citep{Boulares1990,Pfrommer2017b}, implying that they too can influence galactic evolution. Indeed, in \citet{buck2020}, who also used the Auriga model, cosmic rays were found to be able to significantly affect circumgalactic medium properties; altering the angular momentum distribution of the gas and the subsequent development of the stellar disc. Including cosmic ray physics in our simulations could therefore strongly affect the remnant morphology once again. Apart from increasing physical fidelity, re-running our simulations with cosmic ray physics would also allow us to directly examine whether the equiparitition condition holds throughout the mergers. This could naturally have important consequences for inferences made from synchrotron emission observations in the future.

\section{Conclusions}
\label{sec:conclusions}

In this paper, we have investigated the impact of magnetic fields on galaxy mergers. We have done this by comparing MHD and hydrodynamic simulations run from the same initial conditions. Our simulations were fully cosmologically-consistent and used a state-of-the-art zoom-in code. This allowed us to simulate the galaxy and its immediate neighbourhood to high-resolution, whilst still accounting for the influence of large-scale structure. To the best of our knowledge, this is the first time that MHD zoom-in simulations have been used in this way to study galaxy mergers. 

The impact of magnetic fields on mergers is an extremely complex problem and an accurate implementation of the physics involved is technically challenging. In order to increase the reliability of our results, we therefore built upon previously proven work. In particular, our simulations employed the Auriga galaxy formation model and were run using the \textsc{arepo} moving-mesh code. The Auriga model includes a range of physically-motivated subgrid models, with parameters that do not require retuning between resolution levels, and includes an MHD implementation that has been shown to sufficiently fulfil the divergence constraint even in dynamic environments. \textsc{Arepo}, meanwhile, has been shown to have significantly better numerical accuracy than competing codes when applied to a range of relevant physical problems, including several that have direct relevance to our investigation

In total, we ran eight high-resolution simulations, with a dark matter mass resolution $\sim38.5$ times finer than the fiducial Illustris run. We have supported these with two intermediate, and two lower-resolution runs. The global properties of each simulation are broadly converged between resolution level, indicating that the results of our simulations depend on the physics included and not on the numerical implementation. We list our major conclusions from this work below:

\begin{itemize}
    \item Structural properties of the merger remnant depend strongly on the physics model. In particular, MHD simulations produce large disc galaxies that display extensive spiral structure. This is underlain by a flocculent gas disc with a shallow radial gradient. Merger remnants from hydrodynamic simulations, on the other hand, are systematically smaller, and often display distinctive stellar bar and ring elements. The gas discs in these remnants are significantly thinner, with a flatter density profile that drops abruptly at the disc edge (Figs.~\ref{fig:gas}, \ref{fig:mocks}, \ref{fig:profiles}).
    \item Simulations of galaxies with more quiescent merger histories show similar, but less marked, morphological differences. As these galaxies are not completely isolated from tidal interactions during their lifetime, we argue that the morphological differences we observe are predominantly a result of MHD effects excited by mergers (Fig.~\ref{fig:mocks-auriga}).
    \item Global properties are not significantly affected by the inclusion of MHD physics. In particular, the star formation history of the remnant is left relatively unchanged (Fig.~\ref{fig:SFR_dist}). Indeed, rather than suppressing star formation, the total stellar mass is often marginally higher in the MHD runs at $z=0$ (Table~\ref{tab:sim_data},  Fig.~\ref{fig:bound_gas_stellar}). 
    \item Merger remnants in the hydrodynamic simulations lose significantly more gas than than their MHD counterparts (Table~\ref{tab:sim_data},  Fig.~\ref{fig:bound_gas_stellar}). In fact, in some MHD simulations, the loss of gas mass can be accounted for almost entirely by the increased stellar mass. This indicates that we do not see magnetically-driven winds in our simulations. The reduced gas mass in hydrodynamic simulations is likely a result of stronger galactic winds, caused by a similar amount of star formation taking place in a more compact volume (Figs.~\ref{fig:gas} and \ref{fig:mocks}).
    \item During the merger, the mean magnetic field strength in the inner $\lesssim5$ kpc of the disc is boosted by up to an order of magnitude. At the same time, the field outside this range drops by an order of magnitude. The magnetic field can become locally dominant during this time, with the magnetic energy density able to reach several times the thermal energy density in a substantial number of gas cells. These effects are typically apparent for at least 1.5 Gyr after first periapsis and fade with the rebuilding of the disc. (Fig.~\ref{fig:BH_mag}).    
    \item The differences noted above are only observed once sufficient resolution is reached (Figs.~\ref{fig:gas_zoom}, \ref{fig:mocks-zoom}, \ref{fig:profiles-zoom}). This is explained by Fig.~\ref{fig:mag_zoom}, where it is shown that the merger amplifies the galactic magnetic field more strongly in higher resolution simulations. We interpret this as evidence that our results are dependent on the sufficient excitement of a small-scale dynamo (Fig.~\ref{fig:power_spec}).
\end{itemize}

The above results apply particularly to gas-rich major mergers, where the merger remnant is able to re-form a substantial stellar disc. The impact of such mergers on galaxy evolution taken as a whole, however, is likely to be substantial. Furthermore, the influence of magnetic fields only increases when we consider their impact on the stability of gas flows and their impact on important anisotropic processes, such as the transport of cosmic rays. Our results are therefore a clear indication that the inclusion of MHD physics is critical for reliably modelling galaxy evolution in future simulations. A full elucidation of the mechanism behind the morphological differences seen in this paper will be presented in an upcoming work.

\section*{Acknowledgments}
The authors thank the anonymous referee  for a constructive report that  helped improve this paper.
MS and CP acknowledge support by the European Research Council under ERC-CoG grant CRAGSMAN-646955.

\section*{Data Availability}

The data underlying this article will be shared on reasonable request to the corresponding author.

\bibliographystyle{mnras}
\bibliography{main}

\begin{appendix}
\section[Appendix A: Impact of MHD on the ISM]{Impact of MHD on the ISM}
\label{appendix:ISM}

\begin{figure}
    \centering
    \includegraphics[width=\columnwidth]{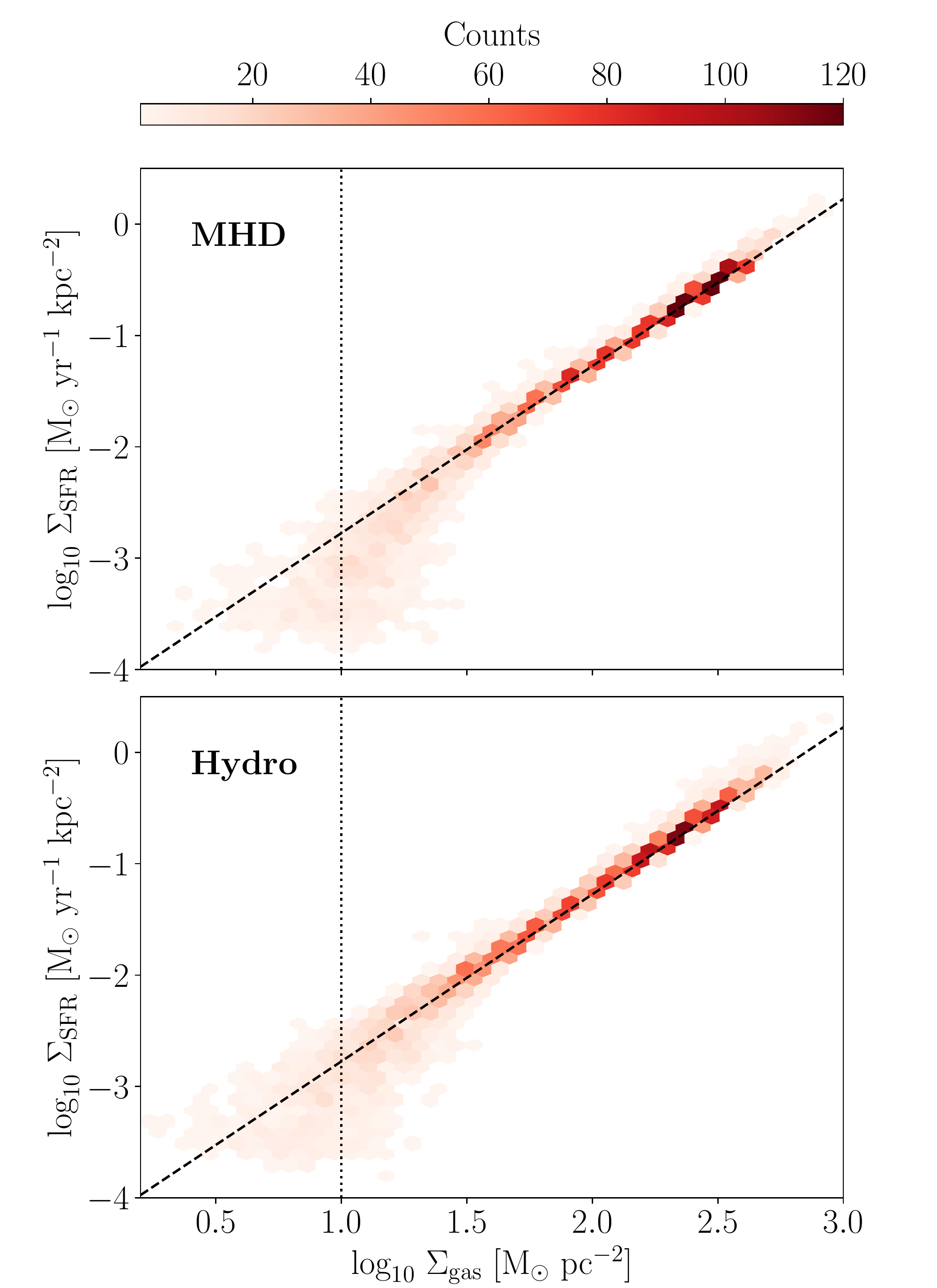}
    \caption{\textit{Top panel:} star formation rate surface density as a function of gas surface density for 1349-3M, as seen at a lookback time of $\sim$4 Gyr. The dashed line shows a Kennicutt-Schmidt relation \citep{schmidt1959, kennicutt1998} with exponent 1.5. The dotted line indicates the approximate position of the cut-off in the star formation rate. \textit{Bottom panel:} as above, but for 1349-3H. Both follow the same relation, despite the strong amplification of the magnetic field in 1349-3M during this time.}
    \label{fig:kennicutt_schmidt}
\end{figure}

In Section~\ref{subsec:set-up}, we claimed that it was not necessary to recalibrate our ISM subgrid model when including magnetic fields in the simulation. This statement is non-trivial: whilst the effective pressure in our ISM subgrid model is a function of density only in the hydrodynamic simulations \citep{springel2003}, in the MHD simulations there is an additional magnetic pressure term to consider. We can check the influence of this additional term on the ISM model by investigating its impact on the relation between the star formation rate surface density, $\Sigma_\text{SFR}$, and the gas surface density, $\Sigma_\text{gas}$. Stars form probabilistically out of our simulated ISM with a gas consumption time-scale set to match that observed by \cite{kennicutt1998} for disc galaxies in the local Universe. This results in the ISM following the well-known Kennicutt-Schmidt \citep{schmidt1959, kennicutt1998} relation of: $\Sigma_\text{SFR} \propto (\Sigma_\text{gas})^n$. If our ISM subgrid model required recalibration, the form of this relation would be dependent on the physics model used.

We show the relation between $\Sigma_\text{SFR}$ and $\Sigma_\text{gas}$ for 1349-3M and 1349-3H in Fig.~\ref{fig:kennicutt_schmidt}. The 1349 simulations were chosen due to the particularly strong amplification in 1349-3M, but naturally the results apply to all our simulations. We have chosen a time when the magnetic field is highly amplified in the MHD simulation. At later times we observe a similar relation, but with the gas density covering a much narrower range. In both cases, the surface densities are calculated by taking a face-on projection with depth $\pm$1 kpc from the midplane. We choose this depth to make sure that we are predominantly considering gas in the disc. It can be seen that for both physics models, the star formation rate surface density follows the Kennicutt-Schmidt relation with exponent $n=1.5$ \citep[cf.][]{springel2003}. This is true over a broad range of values. At lower gas surface density values, the relation becomes more scattered as the star formation threshold density is approached. The peak of the distribution is higher in the MHD simulation than the hydrodynamic analogue, which is a result of morphological differences between the two remnants. The strong similarity between both relations otherwise supports our assertion that the ISM subgrid model must not be recalibrated when introducing magnetic fields.

\section[Appendix B: Galaxy Tracking and Black Hole Dynamics]{Galaxy Tracking and Black Hole Dynamics}
\label{appendix:galaxy_tracking}

\begin{figure*}
    \centering
    \includegraphics[width=\textwidth]{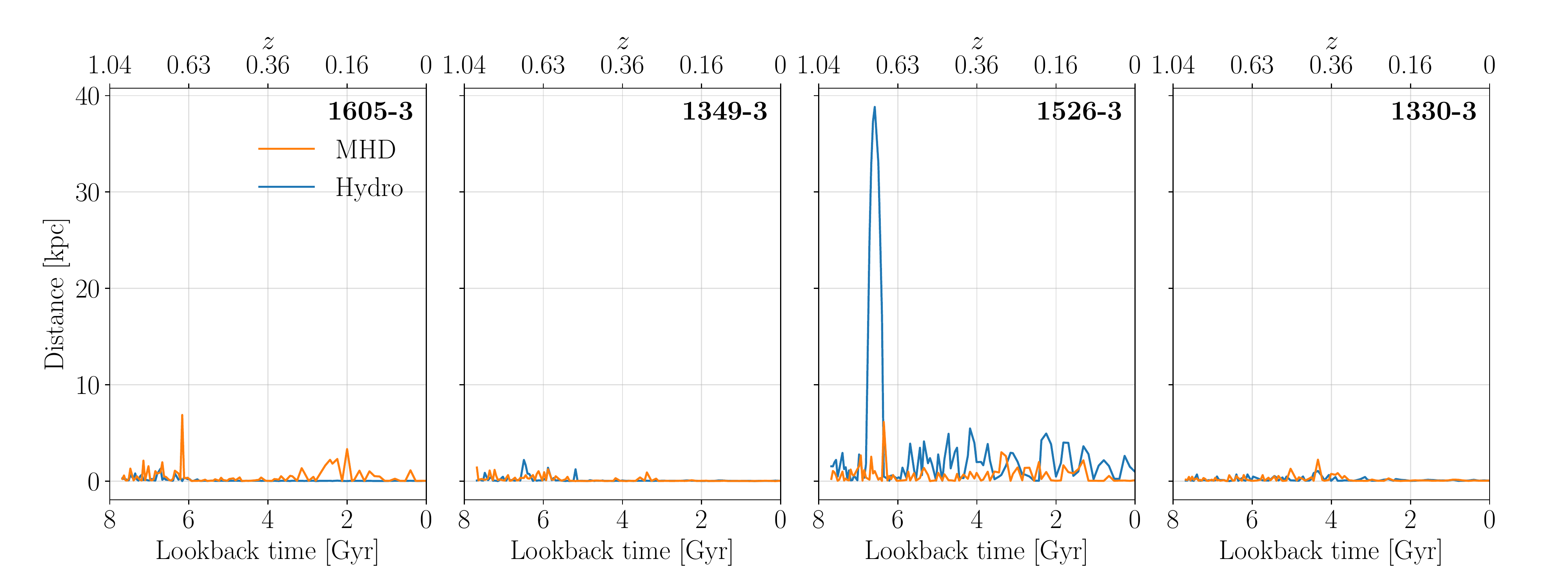}
    \caption{The distance between the galactic centre and the closest black hole for all high-resolution simulations as a function of time. In general, this distance stays well below 5 kpc, confirming the reliability of our galaxy tracking method. The `wandering' nature of the black hole in simulation 1526-3H likely contributes to the unusual morphology displayed by the corresponding merger remnant.}
    \label{fig:black_hole_dist}
\end{figure*}

In Section~\ref{subsec:galaxy_tracking}, we claimed that tracking a galaxy between snapshots is frequently akin to tracking the black hole particle that resides in that galaxy. We support this claim in Fig.~\ref{fig:black_hole_dist}, where we show the distance between the most bound gas cell in the galaxy (i.e. the galactic centre) and the closest black hole for each high-resolution simulation. For the vast majority of snapshots, this distance is always well under 5 kpc, confirming the validity of our tracking method. An exception to this rule is simulation 1526-3H, where the black hole is not well-tied to the galactic centre. In this merger scenario, the secondary progenitor passes directly through the primary progenitor. For a short time immediately afterwards, the black hole then `hitches a ride', becoming gravitationally bound to the merging galaxy. This can be confirmed by comparing its distance from the main galaxy between 7.11 and 6.35 Gyr to the distance between the merging galaxies, as seen in Fig.~\ref{fig:SFR_dist}. 

On the black hole's return to the main galaxy, it never quite loses its newly-gained orbital energy, oscillating around the galactic centre instead. During this time, the black hole continues to accrete gas, and consequently continues to inject energy into neighbouring gas cells. The subsequent AGN outbursts disrupt the gas, producing a similar morphology to that of 1605-3M (see Fig.~\ref{fig:gas}). This similarity is unexpected as the black hole in 1526-3H grows only a quarter as large as that in 1605-3H post-merger, meaning that the AGN outbursts should be significantly less influential (see Section~\ref{subsec:set-up}). Indeed, the black hole growth in 1526-3H is similar to that of 1349-3H and 1330-3H, both of which show no signatures of strong outbursts in their gas morphology. We therefore argue that it is the unlocalised nature of the feedback in 1526-3H that is behind the strongly disrupted morphology. This, in turn, produces the unusual stellar morphology seen in this simulation.

\section[Appendix C: Evidence for a Small-scale Dynamo in Action]{Evidence for a Small-scale Dynamo in Action}
\label{appendix:small-scale_dynamo}

\begin{figure*}
    \centering
    \includegraphics[width=\textwidth]{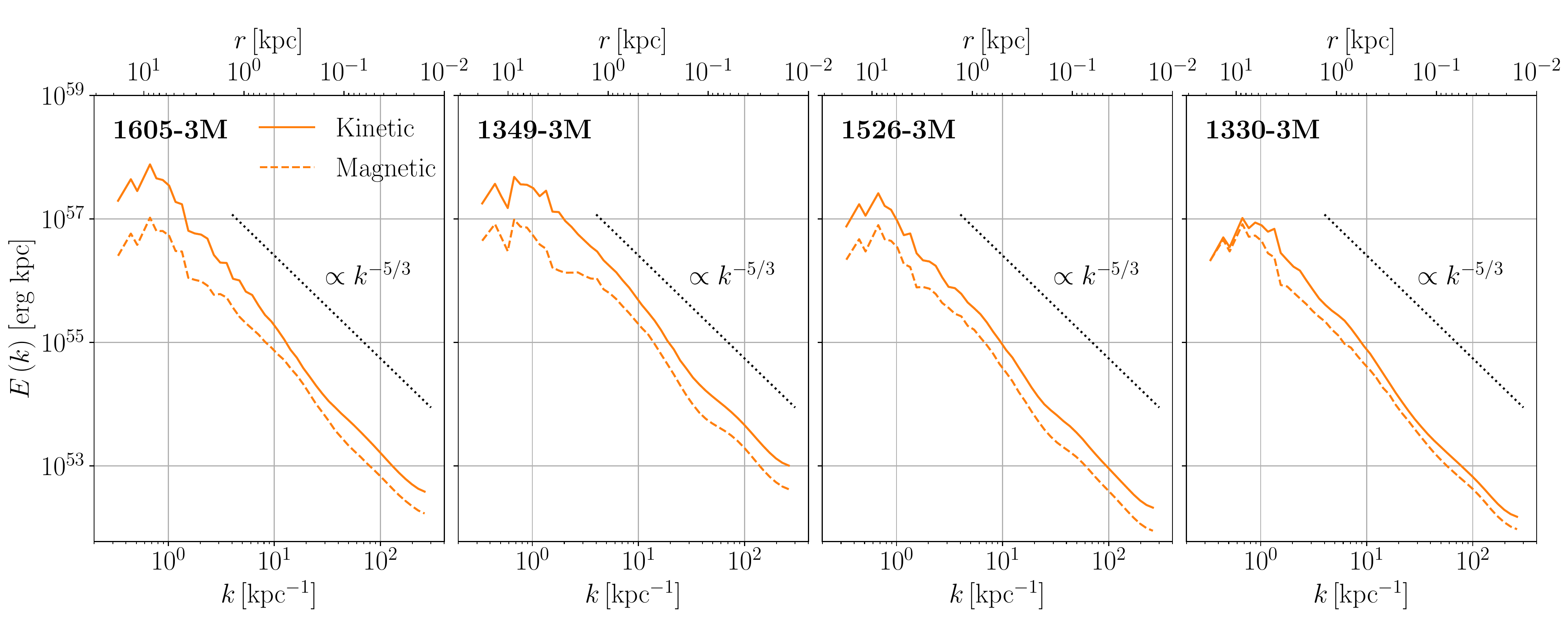}
    \caption{Kinetic and magnetic energy power spectra for the highest-resolution MHD simulations, calculated for gas within a sphere of 5 kpc centred on the galactic centre. The power spectra are shown at 0.5 Gyr after the time of periapsis for each simulation. The black dotted lines show the slopes of a Kolmogorov spectrum ($\propto k^{-5/3}$) \citep{kolmogorov1941}, which is theoretically expected for a small-scale dynamo resulting from incompressible turbulence. In each case, the magnetic field is strongly saturated.}
    \label{fig:power_spec-high-res}
\end{figure*}

In Section~\ref{subsec:amplification_res}, we discussed the development of a small-scale dynamo in our simulations as a function of increasing resolution. In Fig.~\ref{fig:power_spec-high-res}, we support this with kinetic and magnetic energy power spectra for each of our high-resolution MHD simulations. These were created in the same manner as for Fig.~\ref{fig:power_spec} over the same radius of 5 kpc within a zero-padded box of 10 kpc across. The profiles are shown at 0.5 Gyr after the respective time of periapsis in each simulation (see Section~\ref{subsec:sims}). In choosing this time, we show the power spectra at a similar point in the evolution of the magnetic field. It can be seen that for each simulation, the kinetic energy exhibits a Kolmogorov-like spectrum, which is consistent with the volume-filling phase of the gas being both turbulent and subsonic. The difference in normalisation in each plot can be mostly explained by the difference in mass evolution. At high $k$ values, the magnetic energy saturates at around 50 per cent of the kinetic energy, which is consistent with subsonic turbulent box simulations where the forcing was solenoidal \citep{federrath2016}. In each simulation, the magnetic field has grown similarly quickly. Indeed, the magnetic energy is almost fully saturated in each case, with the peak magnetic energy occuring at the driving scale of the turbulence. This means almost the entire magnetic energy spectrum is in the non-linear dynamo phase, and consequently there is little evidence of the Kazantsev-like slope at large scales. The decline of the power spectra at scales larger than 10 kpc is an artefact of the zero-padding we use. Overall, Fig.~\ref{fig:power_spec-high-res} supports our understanding of the growth of the magnetic field, as analysed in Section~\ref{subsec:amplification_res}.

\end{appendix}

\bsp
\label{lastpage}
\end{document}